\documentclass[letterpaper,10pt,twoside]{article}
\usepackage[utf8]{inputenc}
\usepackage[english]{babel}
\usepackage[T1]{fontenc}
\usepackage{lmodern}
\usepackage{amsmath,amssymb,mathtools} 
\usepackage[dvipdfmx]{hyperref}

\newcommand{\N}{\mathbb{N}}
\newcommand{\Z}{\mathbb{Z}}
\newcommand{\R}{\mathbb{R}}
\newcommand{\C}{\mathbb{C}}
\newcommand{\defeq}{\coloneqq}
\newcommand{\tens}{\otimes}
\DeclareMathOperator{\ctens}{\hat{\otimes}}
\newcommand{\ds}{\circ}
\newcommand{\cH}{\mathcal{H}}
\newcommand{\toi}{\hookrightarrow}

\newcommand{\sig}[1]{[#1]}
\newcommand{\fdg}[1]{|#1|}

\newcommand{\cS}{\mathcal{S}}
\newcommand{\cA}{\mathcal{A}}

\newcommand{\hsl}{\langle}
\newcommand{\hsr}{\rangle^{\mathrm{HS}}}
\newcommand{\cB}{\mathcal{B}}

\newcommand{\cop}{\cB}
\newcommand{\copd}{\cB^\ds}

\newcommand{\coprd}{\cB^{\R\ds}}
\newcommand{\copp}{\cB^{+}}
\newcommand{\coppd}{\cB^{+\ds}}
\newcommand{\cophs}{\tilde{\mathcal{B}}}
\newcommand{\cophsd}{\cophs^{\ds}}
\newcommand{\cophsr}{\cophs^{\R}}
\newcommand{\cophsrd}{\cophs^{\R\ds}}
\newcommand{\cophsp}{\cophs^{+}}
\newcommand{\cophspd}{\cophs^{+\ds}}
\newcommand{\po}{\mathsf{P}}
\newcommand{\lhs}{\langle\langle}
\newcommand{\rhs}{\rangle\rangle}
\newcommand{\cD}{\mathcal{D}}
\newcommand{\cDd}{\mathcal{D}^{\ds}}
\newcommand{\cDr}{\mathcal{D}^{\R}}
\newcommand{\cDrd}{\mathcal{D}^{\R\ds}}
\newcommand{\cDp}{\mathcal{D}^{+}}
\newcommand{\cDpd}{\mathcal{D}^{+\ds}}
\newcommand{\obs}{\mathcal{O}}
\newcommand{\pobs}{\mathcal{E}}
\newcommand{\aglue}{\diamond}

\allowdisplaybreaks

\begin{document}

\begin{titlepage}
\title{\textbf{A positive formalism\\ for quantum theory\\ in the general boundary formulation}}
\author{Robert Oeckl\footnote{email: robert@matmor.unam.mx}\\ \\
Centro de Ciencias Matemáticas,\\
Universidad Nacional Autónoma de México,\\
C.P.~58190, Morelia, Michoacán, Mexico}
\date{UNAM-CCM-2012-5\\ 21 December 2012\\ 30 August 2013 (v2)}

\maketitle

\vspace{\stretch{1}}

\begin{abstract}
We introduce a new ``positive formalism'' for encoding quantum theories in the general boundary formulation, somewhat analogous to the mixed state formalism of the standard formulation. This makes the probability interpretation more natural and elegant, eliminates operationally irrelevant structure and opens the general boundary formulation to quantum information theory.

\end{abstract}

\vspace{\stretch{1}}
\end{titlepage}



\section{Introduction}

The standard formulation of quantum theory (as laid out for example in von~Neumann's book \cite{vne:mathgrundquant}) relies on a fixed a priori notion of time. While this is unnatural from a special relativistic perspective, it is not irreconcilable, as shown by the success of quantum field theory. It flatly contradicts general relativistic principles, however. This has been a key difficulty in bringing together quantum theory and general relativity.

In contrast, the general boundary formulation (GBF) of quantum theory \cite{Oe:gbqft}, relies merely on a weak (topological) notion of spacetime and is thus compatible from the outset with general relativistic principles. The GBF is indeed motivated by the problem of providing a suitable foundation for a quantum theory of gravity \cite{Oe:catandclock}. However, it is also motivated by the stunning empirical success of quantum field theory. In particular, the GBF is an attempt to learn about the foundations of quantum theory from quantum field theory \cite{Oe:reveng}.

So far, the GBF has been axiomatized in analogy to the pure state formalism of the standard formulation. In particular, Hilbert spaces are basic ingredients of the formalism in both cases.\footnote{In the presence of fermionic degrees of freedom the GBF requires the slight generalization from Hilbert spaces to Krein spaces \cite{Oe:freefermi}.} The elements of these Hilbert spaces have been termed ``states'' in the GBF as in the standard formulation, although they do not necessarily have the same interpretation. In fact, projection operators play a more fundamental role in the probability interpretation of the GBF than ``states''. This suggests to construct a formalism for the GBF where this fundamental role is reflected mathematically. This is somewhat analogous to the transition from a pure state to a mixed state formalism in the standard formulation, where in the latter also (positive normalized trace-class) operators play a more fundamental role.

There are various motivations for introducing such a formalism. One is operationalism. As already alluded to, the new formalism is intended to bring to the forefront the objects of more direct physical relevance, eliminating operationally irrelevant structure and information. It turns out that this leads to a more simple and elegant form of the probability interpretation. At the same time, the positivity of probabilities becomes imprinted on the formalism in a rather direct way, via order structures on vector spaces. Therefore we term the new formalism the \emph{positive formalism}. In contrast we shall refer to the usual formalism as the \emph{amplitude formalism}. A consequence of the elimination of superfluous structure is a corresponding widening of the concept of a quantum theory. We expect this to be beneficial both in understanding quantum field theories from a GBF perspective as well as in the construction of completely new theories, including approaches to quantum gravity. Another motivation is the opening of the GBF to quantum information theory. In particular, the positive formalism should facilitate the implementation of general quantum operations in the GBF as well as the introduction of information theoretic concepts such as entropy.

In Section~\ref{sec:mixedsf} we expand on the motivation for the positive formalism by drawing on the analogy to the mixed state formalism in the standard formulation. In Section~\ref{sec:bosonic} we introduce the positive formalism for purely bosonic quantum theory. This restriction makes it simpler and conceptually more transparent than the general case. The first step in this is a review of the probability interpretation and expectation values in Section~\ref{sec:bprob}. This leads to the definition of two new objects of central importance to the positive formalism, the \emph{probability map} and the \emph{expectation map}. Elementary properties of the former are exhibited in Section~\ref{sec:probprop}. The compatibility of the probability maps with the notion of composition from spacetime gluing is demonstrated in Section~\ref{sec:bcomp}. In Section~\ref{sec:bhsurfaces} the appropriate accompanying data for hypersurfaces is determined, culminating in Section~\ref{sec:bpaxioms} in the proposal of a complete axiomatic system for the resulting positive formalism. A corresponding axiomatization of observables is considered in Section~\ref{sec:bobs}. The positive formalism for the general case, including fermionic degrees of freedom, is developed in Section~\ref{sec:fermionic}. Again, the first step is a discussion of the probability interpretation, in this case limited to the modifications imposed by the fermionic grading and Krein space structure. Realness and positivity of the relevant structures is treated in Section~\ref{sec:frps} and the compatibility with composition in Section~\ref{sec:fcomp}. This leads to the axiomatization of the positive formalism in the general case in Section~\ref{sec:fpaxioms}. Observables are considered in Section~\ref{sec:fobs}. In Section~\ref{sec:outlook} we offer a discussion of the results from various perspectives, indicate some open questions and point to some directions of future research. In order to make this article reasonably self-contained the axioms of the amplitude formalism of the GBF are recalled in Appendix~\ref{sec:gbfaxioms}.



\section{Motivation: Mixed states and positivity}
\label{sec:mixedsf}

In order to describe a quantum system statistically in the standard formulation we need to admit ensembles of (pure) states. Recall that the standard way to encode ensembles is via density operators \cite{vne:mathgrundquant}. Say, we consider a family of states $\{\psi_i\}_{i\in I}$ in the separable Hilbert space $\cH$ of the system, indexed by a finite or countably infinite set $I$. We associate probabilities $\{p_i\}_{i\in I}$ to the states of the ensemble. The latter satisfy $p_i\in [0,1]$ for all $i\in I$ and sum to unity, i.e., $\sum_{i\in I} p_i=1$. This ensemble or \emph{mixed state} is encoded via the operator $\sigma$ on $\cH$ given by
\begin{equation}
 \sigma=\sum_{i\in I} p_i \po_i,
\label{eq:tcnop}
\end{equation}
where $\po_i$ is the orthogonal projector onto the subspace spanned by $\psi_i$. Note that two ensembles defined in this way are physically equivalent if the operators associated to them via (\ref{eq:tcnop}) are identical. The density operators, i.e., the operators of the form (\ref{eq:tcnop}) are precisely the positive operators of unit trace on $\cH$.

The Hilbert-Schmidt inner product yields a symmetric pairing between mixed states,
\begin{equation}
\hsl \sigma_2,\sigma_1\hsr =\sum_{n\in N} \langle \sigma_2\xi_n,\sigma_1\xi_n\rangle .
\label{eq:msip}
\end{equation}
Here $\{\xi_n\}_{n\in N}$ denotes an orthonormal basis of $\cH$.
If $\sigma_1$ and $\sigma_2$ encode pure states, i.e., are one-dimensional orthogonal projectors, then $\hsl \sigma_2,\sigma_1\hsr$ is precisely the probability of instantaneously measuring $\sigma_2$ given that $\sigma_1$ was prepared.\footnote{A notion of probability for measuring a general mixed state given that another was prepared can also be defined, but is more involved \cite{Uhl:transprobstar}.}

Suppose the system evolves in time. Say the unitary operator $U$ on $\cH$ describes the evolution from an initial time $t_1$ to a final time $t_2$. The evolution of the mixed state is simply given by the conjugation operator $\tilde{U}:\cop\to\cop$ with
\begin{equation}
\tilde{U}(\sigma)\defeq U\sigma U^{-1} .
\end{equation}
Consider two consecutive time evolutions, first from $t_1$ to $t_2$ and then from $t_2$ to $t_3$. Then the operator $U_{[t_1,t_3]}$ for evolution directly from $t_1$ to $t_3$ is of course the composition of the operators $U_{[t_1,t_2]}$ and $U_{[t_2,t_3]}$ for the individual time evolutions,
\begin{equation}
 U_{[t_1,t_3]}=U_{[t_2,t_3]}\circ U_{[t_1,t_2]} .
\end{equation}
This property is directly inherited by the corresponding operators $\tilde{U}$,
\begin{equation}
 \tilde{U}_{[t_1,t_3]}=\tilde{U}_{[t_2,t_3]}\circ \tilde{U}_{[t_1,t_2]} .
\end{equation}

Apart from the possibility to describe ensembles, this way of encoding states has the interesting property of containing less information as compared to the pure state formalism. For example, the ensemble $\{\psi_i,p_i\}_{i\in I}$ as considered above is not uniquely determined by the operator (\ref{eq:tcnop}), even if we limit ourselves to pure states. This loss of information, albeit small here, should be considered welcome as the information in question (essentially a phase factor) is physically irrelevant.

A related aspect of the mixed state formalism is that it brings real structure, positivity properties and thus order structure to the foreground as properties of probability.\footnote{Recall that an ordered vector space is a real vector space equipped with a compatible partial order relation. Compatible means that the order relation is invariant under translations and under scalar multiplication with positive numbers. The order relation is completely determined by the set of positive elements, i.e., the elements that are larger than or equal to $0$.} Concretely, the Hilbert-Schmidt inner product $\hsl \cdot,\cdot\hsr$ is real on the real vector space of self-adjoint Hilbert-Schmidt operators making the latter into a real Hilbert space. The subset of positive Hilbert-Schmidt operators induces a partial order making it into an ordered vector space. Restricting the inner product to these positive operators makes it also positive, as used in the interpretation of (\ref{eq:msip}) as a probability. The time-evolution operator $\tilde{U}$ restricted to self-adjoint operators produces self-adjoint operators. Moreover, it is positive, i.e., it maps positive operators to positive operators. It also conserves the trace so that it maps mixed states to mixed states. These considerations suggest that positivity and order structure should play a more prominent role at a foundational level than say the Hilbert space structure of $\cH$ or the algebra structure of the operators on it from which they are usually derived.

Algebraic quantum field theory \cite{Haa:lqp} is a great example of the fruitfulness of taking serious some of these issues. There, one abandons in fact the notion of Hilbert spaces in favor of more flexible structures build on $C^*$-algebras. Also, positivity plays a crucial role there in the concept of state.



\section{Bosonic theory}
\label{sec:bosonic}

A \emph{positive formalism} somewhat analogous to the mixed state formalism of the standard formulation can be introduced in the GBF. We shall develop this in the present paper. For the reader's convenience the usual axioms of the GBF are recalled in Appendix~\ref{sec:gbfaxioms}. This includes the notion of spacetime employed in the GBF (Appendix~\ref{sec:sts}), the core axioms (Appendix~\ref{sec:coreaxioms}) and the observable axioms (Appendix~\ref{sec:obsax}). We restrict at first to the case of purely bosonic quantum theory as this turns out to be simpler and more intuitive than the general case. 

\subsection{Probabilities and expectation values}
\label{sec:bprob}

In the present section we exhibit the local ingredients of the positive formalism and their role in the description of measurements. We shall see that this lends a particular elegance and compelling simplicity to the probability interpretation of the GBF.

Consider a bosonic general boundary quantum field theory, i.e., a quantum theory in terms of the bosonic axioms of the GBF \cite{Oe:gbqft}. (The general axioms are also listed in Appendix~\ref{sec:coreaxioms}. In the bosonic case the spaces $\cH_{\Sigma}$ are purely even in f-degree and may be taken to be Hilbert spaces \cite{Oe:freefermi}.) Given a hypersurface $\Sigma$ and associated Hilbert space $\cH_{\Sigma}$ we denote by $\cop_{\Sigma}$ the algebra of continuous operators on $\cH_{\Sigma}$.

We start with probabilities associated to measurements located on boundaries of spacetime regions \cite{Oe:gbqft}. Given a spacetime region $M$ that comprises the system of interest we choose two closed subspaces $\cS,\cA$ of the boundary Hilbert space $\cH_{\partial M}$ such that $\cA\subseteq \cS\subseteq\cH_{\partial M}^{\ds}$. (Recall that $\cH_{\partial M}^{\ds}$ is the dense subspace of $\cH_{\partial M}$ where the amplitude map $\rho_M$ is defined, see Axiom (T4) in Appendix~\ref{sec:coreaxioms}.) The subspace $\cS$ encodes knowledge we have about the measurement such as a preparation that we have performed. The subspace $\cA$ encodes a question about the system. We then denote by $P(\cA|\cS)$ the probability for measuring an affirmative answer. We might intuitively describe this as the probability that the system is found to ``be'' in the subspace $\cA$ given that we know it to ``be'' in the subspace $\cS$.

Denote by $\{\xi_n\}_{n\in N}$ an orthonormal basis of $\cH_{\partial M}$ in $\cH_{\partial M}^{\ds}$, where $N$ is finite or $N=\N$. Suppose $N_{\cA}$ and $N_{\cS}$ are subsets of $N$ with $N_{\cA}\subseteq N_{\cS}\subseteq N$, indexing orthonormal basis of $\cA$ and $\cS$ respectively. The probability $P(\cA|\cS)$ is given by the formula,\footnote{Expression (\ref{eq:aprob}) may be ill defined due to numerator or denominator being infinite or due to the denominator being zero. We shall assume that this is not the case. This implicit restriction on $\cS$ and $\cA$ is not immediately relevant to the following considerations.}
\begin{equation}
 P(\cA|\cS)=\frac{\sum_{n\in N_{\cA}} |\rho_M(\xi_n)|^2}{\sum_{n\in N_{\cS}} |\rho_M(\xi_n)|^2} .
\label{eq:aprob}
\end{equation}
While numerator and denominator are quadratic expressions here, we may rewrite them in terms of a linear dependence on the operators $\po_{\cA}$ and $\po_{\cS}$, which project onto the subspaces $\cA$ and $\cS$ respectively, as follows,
\begin{equation}
 P(\cA|\cS)=\frac{\sum_{n\in N} \overline{\rho_M(\xi_n)}\rho_M(\po_A\xi_n)}{\sum_{n\in N} \overline{\rho_M(\xi_n)}\rho_M(\po_S\xi_n)} .
\label{eq:aprobl}
\end{equation}
This suggests to define the complex linear map $A_M:\copd_{\partial M}\to\C$ which we shall refer to as the \emph{probability map},
\begin{equation}
A_M(\sigma)\defeq\sum_{n\in N} \overline{\rho_M(\xi_n)}\rho_M(\sigma \xi_n) .
\label{eq:defpm}
\end{equation}
Here $\copd_{\partial M}$ is a suitable subspace of the space $\cop_{\partial M}$ of continuous operators on the boundary Hilbert space $\cH_{\partial M}$.\footnote{The precise determination of $\copd_{\partial M}$ is not relevant here and will be discussed later.}
This definition renders the probability formula (\ref{eq:aprob}) remarkably simple,
\begin{equation}
 P(\cA|\cS)=\frac{A_M(\po_{\cA})}{A_M(\po_{\cS})} .
\label{eq:nprob}
\end{equation}

Observe that expression (\ref{eq:nprob}) is linear in the operator $\po_{\cA}$ which encodes the ``question''. This allows to implement an ensemble of alternative questions with associated weights in a direct way. Say, instead of asking for $\cA$ we ask with weights $a_1,\dots,a_n$ for $\cA_1,\dots,\cA_n$ (where $\cA_i\subseteq \cS$). That is, we are asking for an \emph{expectation value}. This is given by,
\begin{equation}
 \sum_{i=1}^n a_i P(\cA_i|\cS)
 =\sum_{i=1}^n a_i \frac{A_M(\po_{\cA_i})}{A_M(\po_{\cS})}
 = \frac{A_M(Q)}{A_M(\po_{\cS})},
\label{eq:boexpval}
\end{equation}
where we have defined the operator,
\begin{equation}
 Q\defeq \sum_{i=1}^n a_i \po_{\cA_i} .
\end{equation}
If we impose the conditions $0<a_i\le 1$ and $a_1+\cdots+a_n=1$, the quantities $a_i$ may be interpreted as probabilities. $Q$ is then a positive operator that we may think of as encoding an ensemble of quantum boundary conditions or an ensemble of measurements. On the other hand, without any constraint on the quantities $a_i$, the expression (\ref{eq:boexpval}) is still a kind of expectation value. Indeed, it is then largely analogous to the notion of expectation value in the standard formulation and we may say that $Q$ encodes a \emph{boundary observable}.

Recall, however, that the observable concept in the GBF is more general \cite{Oe:obsgbf}. In particular, observables may be associated to spacetime regions. Thus, an observable in the spacetime region $M$ is encoded in an \emph{observable map} $\rho_M^O:\cH_{\partial M}^{\ds}\to\C$. As for simple probabilities, a subspace $\cS\subseteq \cH_{\partial M}^{\ds}$ encodes knowledge about the measurement process such as a preparation. The expectation value of the observable encoded by $\rho_M^O$ is then given by the formula,
\begin{equation}
 \langle O \rangle_{\cS}=\frac{\sum_{n\in N}\overline{\rho_M(\xi_n)}\rho_M^O(\po_{\cS}\xi_n)}{\sum_{n\in N} \overline{\rho_M(\xi_n)}\rho_M(\po_S\xi_n)} .
\label{eq:expval}
\end{equation}
A boundary observable and its expectation value are recovered in the special case when the observable map arises as the composition of the amplitude map with the operator encoding the boundary observable, $\rho_M^O=\rho_M\circ Q$.
We have written (\ref{eq:expval}) in a way to emphasize the similarity with (\ref{eq:aprobl}). This suggests to define the complex linear map $A_M^O:\copd_M\to\C$ in close analogy to (\ref{eq:defpm}),
\begin{equation}
A_M^O(\sigma)\defeq\sum_{n\in N} \overline{\rho_M(\xi_n)}\rho_M^O(\sigma \xi_n) .
\label{eq:defexpm}
\end{equation}
We shall refer to $A_M^O$ as the \emph{expectation map}.
We obtain the compellingly simple formula for the expectation value,
\begin{equation}
\langle O \rangle_{\cS}=\frac{A_M^O(\po_{\cS})}{A_M(\po_{\cS})} .
\label{eq:nexpval}
\end{equation}

Apart from making the formulas for probabilities and expectation values more simple and elegant, the transition from amplitude maps to probability maps and from observable maps to expectation maps has further important implications.
 For one, the latter contain slightly less information than the former. Similar to the case of the mixed state formalism in the standard formulation we lose a physically irrelevant phase.
 Another consequence is the shift in emphasis from the Hilbert space $\cH_{\Sigma}$ to the space $\cop_{\Sigma}$ of operators on it. Superficially, this appears to also be in analogy to the mixed state formalism of the standard formulation, but there are profound differences. In contrast to the standard formulation, the elements of $\cH_{\Sigma}$ do not in general have the interpretation of conventional states. Similarly, the relevant elements of $\cop_{\Sigma}$ do not have the interpretation of ensembles of states. Indeed, as apparent in formulas (\ref{eq:nprob}) and (\ref{eq:nexpval}) relevant operators are projection operators. Rather than representing states or ensembles we can think of them as representing \emph{quantum boundary conditions}. Only in special cases can these be seen to arise from states in the conventional sense.

\subsection{Properties of the probability map}
\label{sec:probprop}

The definition (\ref{eq:defpm}) of the probability map implies its positivity. Denote by $\coppd_M$ the set of positive operators in $\copd_M$. Then, given $\sigma\in\coppd_M$ we have
\begin{equation}
A_M(\sigma)\ge 0 .
\end{equation}
Indeed, this positivity is essential for the probability formula (\ref{eq:nprob}). It ensures non-negativity of numerator and denominator and thus non-negativity of the quotient $P(\cA|\cS)$ (provided $A_M(\po_{\cS})\neq 0$ so it is defined). Also since $\cA\subseteq\cS$ we have $\po_{\cA}\le \po_{\cS}$ and so positivity implies $A_M(\po_{\cA})\le A_M(\po_{\cS})$, guaranteeing that the probability $P(\cA|\cS)$ is less than or equal to one.

Note also that either as a consequence of positivity or by direct inspection of (\ref{eq:defpm}), the probability map is real. That is, given a self-adjoint operator $\sigma\in \coprd_M$ (this is how we denote the real subspace of self-adjoint operators), we have
\begin{equation}
A_M(\sigma)\in\R .
\end{equation}

If $\cH_M$ is infinite-dimensional, positivity also suggests a way of extending the definition of $A_M$ to the whole set of positive operators $\copp_M$ in $\cop_M$. To this end denote the restriction of $A_M$ to $\coppd_M$ by $A_M^+:\coppd_M\to [0,\infty)$. We then extend the range of $A_M^+$ to $[0,\infty]$, seen as the one-point compactification of $[0,\infty)$. As above we choose an orthonormal basis $\{\xi_n\}_{n\in \N}$ of $\cH_{\partial M}$ in $\cH_{\partial M}^{\ds}$. Define $\po_n$ to be the projector onto the subspace spanned by $\{\xi_1,\dots,\xi_n\}$. Given a positive operator $\sigma\in\copp_M$, we define the positive operator $\sigma_n\defeq \po_n\sigma \po_n$. Then, $\sigma_n\in \coppd_M$.\footnote{Even though we have not defined $\copd_M$ before, the definition (\ref{eq:defpm}) and the fact that $\rho_M$ is defined on $\cH_M^{\ds}$ imply that we may take the vector space $\copd_M$ to include all orthogonal projectors onto 1-dimensional subspaces of $\cH_{M}^{\ds}$. This is sufficient for the present purposes.} Moreover, $\{\sigma_n\}_{n\in \N}$ converges to $\sigma$ in the weak operator topology. What is more, the sequence $\{\sigma_n\}_{n\in \N}$ is increasing, i.e., $\sigma_n\ge\sigma_k$ if $n\ge k$. This implies that the following limit
\begin{equation}
A_M(\sigma)\defeq \lim_{n\to\infty} A_M(\sigma_n)
\end{equation}
exists in $[0,\infty]$.


\subsection{Composition}
\label{sec:bcomp}

In Section~\ref{sec:bprob} we have seen that the probability interpretation simplifies when we replace the amplitude map by the probability map. This observation becomes more intriguing when we realize that the probability map behaves well under composition and is thus suitable for replacing the amplitude map at an axiomatic level.

Consider first the disjoint gluing of regions. Assume the context of bosonic Axiom (T5a) of Appendix~\ref{sec:coreaxioms}. For brevity we write $\tau_{\partial M_1,\partial M_2;\partial M}$ as $\tau$. We also define
\begin{equation}
\tau^*:\cop_{\partial M_1}\tens\cop_{\partial M_2}\to\cop_{\partial M}
\quad\text{as}\quad
\tau^*(\sigma_1\tens\sigma_2)\defeq \tau\circ (\sigma_1\tens\sigma_2)\circ\tau^{-1} .
\label{eq:taubos}
\end{equation}
Let $\{\xi_{1,n}\}_{n\in N_1}$ be an orthonormal basis of $\cH_{\partial M_1}$ in $\cH^{\ds}_{\partial M_1}$ and $\{\xi_{2,n}\}_{n\in N_2}$ an orthonormal basis of $\cH_{\partial M_2}$ in $\cH^{\ds}_{\partial M_2}$. Then, $\{\tau(\xi_{1,n}\tens\xi_{2,m})\}_{(n,m)\in N_1\times N_2}$ is an orthonormal basis of $\cH_{\partial M}$ in $\cH^{\ds}_{\partial M}$. Let $\sigma_1\in\copd_{\partial M_1}$ and $\sigma_2\in\copd_{\partial M_2}$. Combining the definition (\ref{eq:defpm}) and the gluing identity (\ref{eq:glueid5a}) yields,
\begin{align}
& A_M\left(\tau^*(\sigma_1\tens\sigma_2)\right) \nonumber \\
& = \sum_{n,m} \overline{\rho_M(\tau(\xi_{1,n}\tens\xi_{2,m}))}\rho_M(\tau(\sigma_1\xi_{1,n}\tens\sigma_2\xi_{2,m})) \nonumber \\
& = \sum_{n,m} \overline{\rho_{M_1}(\xi_{1,n})\rho_{M_2}(\xi_{2,m})}\rho_{M_1}(\sigma_1\xi_{1,n})\rho_{M_2}(\sigma_2\xi_{2,m}) \nonumber \\
& = A_{M_1}(\sigma_1) A_{M_2}(\sigma_2) .
\label{eq:t5atrans}
\end{align}
The structural similarity of the resulting gluing identity for probability maps with the corresponding gluing identity (\ref{eq:glueid5a}) for the amplitude maps is striking.
This identity preserves positivity and realness manifestly in the following sense. Given $\sigma_1$ and $\sigma_2$ positive, their tensor product is positive and so is thus $\tau^*(\sigma_1\tens\sigma_2)$. The positivity of $A_M$ is then induced by the positivity of $A_{M_1}$ and $A_{M_2}$. The analogous statement holds for the realness of $A_M$ with respect to self-adjoint operators.

We proceed to consider the gluing along hypersurfaces. Assume the context of bosonic Axiom (T5b) of Appendix~\ref{sec:coreaxioms}. We write $\tau_{\Sigma_1,\Sigma,\overline{\Sigma'};\partial M}$ as $\tau$ and $c(M;\Sigma,\overline{\Sigma'})$ as $c$. Also, we define $\tau^*:\cop_{\Sigma_1}\tens\cop_{\Sigma}\tens\cop_{\overline{\Sigma'}}\to\cop_{\Sigma}$ and $\iota_{\Sigma}^*:\cop_{\Sigma}\to\cop_{\overline{\Sigma}}$ in the obvious ways.
Let $\{\xi_n\}_{n\in N_1}$ be an orthonormal basis of $\cH_{\Sigma_1}$ in $\cH_{\Sigma_1}^{\ds}$. Likewise, let $\{\zeta_m\}_{m\in N}$ be an orthonormal basis of $\cH_{\Sigma}$ in $\cH_{\Sigma}^{\ds}$. Define the operators $e_{l k}\in\cop_{\Sigma}$ as $e_{l k}\zeta_i\defeq \delta_{k,i}\zeta_l$. Let $\sigma\in\copd_{\partial M_1}$. Combining the definition (\ref{eq:defpm}) and the gluing identity (\ref{eq:glueid5b}) yields,
\begin{align}
& A_{M_1}(\sigma) \cdot |c|^2 \nonumber\\
& = \sum_{n\in N_1} \overline{\rho_{M_1}(\xi_n)\cdot c}\;\rho_{M_1}(\sigma \xi_n)\cdot c \nonumber\\
& = \sum_{n\in N_1; k,l\in N} \overline{\rho_M\left(\tau(\xi_n\tens\zeta_k\tens\iota_{\Sigma}(\zeta_k))\right)}\;
\rho_M\left(\tau(\sigma \xi_n\tens\zeta_l\tens\iota_{\Sigma}(\zeta_l))\right) \nonumber\\
& = \sum_{n\in N_1; k,l\in N} \overline{\rho_M\left(\tau(\xi_n\tens\zeta_k\tens\iota_{\Sigma}(\zeta_k))\right)}\;
\rho_M\left(\tau(\sigma \xi_n\tens e_{l k}\zeta_k\tens\iota_{\Sigma}(e_{l k}\zeta_k))\right) \nonumber\\
& = \sum_{n\in N_1; k,l,i,j\in N} \overline{\rho_M\left(\tau(\xi_n\tens\zeta_i\tens\iota_{\Sigma}(\zeta_j))\right)}\;
\rho_M\left(\tau(\sigma \xi_n\tens e_{l k}\zeta_i\tens\iota_{\Sigma}(e_{l k}\zeta_j))\right) \nonumber\\
& = \sum_{k,l\in N} A_M\left(\tau^*(\sigma\tens e_{l k}\tens \iota_{\Sigma}^*(e_{l k}))\right) .
\label{eq:glueprob1}
\end{align}

Again, the structural similarity of the resulting gluing identity for the probability map to the corresponding identity (\ref{eq:glueid5b}) for the amplitude map is striking. It suggests moreover, to interpret $\{e_{l k}\}_{(l,k)\in N\times N}$ as an orthonormal basis.
Indeed, consider the subalgebra $\cophs_{\partial M}\subseteq\cop_{\partial M}$ of \emph{Hilbert-Schmidt} operators. These are the operators for which the inner product
\begin{equation}
\lhs \sigma_2, \sigma_1\rhs_{\partial M}\defeq \sum_{n\in N}
\langle\sigma_2 \zeta_n,\sigma_1\zeta_n\rangle_{\partial M}
\label{eq:hsip}
\end{equation}
is well defined. This inner product makes $\cophs_{\partial M}$ into a Hilbert space canonically isomorphic to the Hilbert space $\cH_{\partial M}\ctens\cH_{\partial M}^*$. (Here $\ctens$ denotes the completed tensor product.) $\{e_{l k}\}_{(l,k)\in N\times N}$ is an orthonormal basis of $\cophs_{\partial M}$. What is more, as in the corresponding identity for the amplitude map we may replace this orthonormal basis with any other one. Given such an orthonormal basis $\{\eta_m\}_{m\in I}$ of $\cophs_{\partial M}$ we may thus write the resulting identity as,
\begin{equation}
A_{M_1}(\sigma) \cdot |c|^2
= \sum_{m\in I} A_M\left(\tau^*(\sigma\tens \eta_m\tens \iota_{\Sigma}^*(\eta_m))\right) .
\label{eq:bglueid5b}
\end{equation}

Positivity and realness of this identity can be seen as follows. Consider the real subspace $\cophsr_{\partial M}\subseteq\cophs_{\partial M}$ of \emph{self-adjoint} Hilbert-Schmidt operators. The inner product (\ref{eq:hsip}) is real and symmetric on $\cophsr_{\partial M}$ making it into a real Hilbert space. In fact, $\cophs_{\partial M}$ may be seen as the complexification of $\cophsr_{\partial M}$. An orthonormal basis of $\cophsr_{\partial M}$ as a real Hilbert space is also an orthonormal basis of $\cophs_{\partial M}$ as a complex Hilbert space. Picking such a basis the operator $\tau^*(\sigma\tens \eta_m\tens \iota_{\Sigma}^*(\eta_m))$ is self-adjoint if $\sigma$ is self-adjoint, making the identity manifestly real. To establish positivity we have to take into account the summation on the right hand side of the identity. Going back to the form (\ref{eq:glueprob1}), it is easy to verify explicitly that the operator $\sum_{k,l\in N} e_{lk}\tens \iota^*_{\Sigma}(e_{l k})$ is positive on $\cH_{\Sigma}\tens\cH_{\overline{\Sigma}}$. Thus, if $\sigma$ is positive, so is $\sum_{k,l\in N} \tau^*(\sigma\tens e_{l k}\tens \iota_{\Sigma}^*(e_{l k}))$.

Apart from the gluing Axioms (T5a) and (T5b), the amplitude map also enters in Axiom (T3x) which establishes its relation to the inner product of $\cH_{\Sigma}$. Consider thus the context of the bosonic Axiom (T3x) of Appendix~\ref{sec:coreaxioms}. For brevity we write $\tau_{\overline{\Sigma},\Sigma';\partial\hat{\Sigma}}$ as $\tau$. We also define $\tau^*:\cop_{\overline{\Sigma}}\tens\cop_{\Sigma}\to\cop_{\partial\hat{\Sigma}}$ and $\iota_{\Sigma}^*:\cop_{\Sigma}\to\cop_{\overline{\Sigma}}$ in the obvious way. Let $\{\xi_n\}_{n\in N}$ be an orthonormal basis of $\cH_{\Sigma}$ in $\cH_{\Sigma}^{\ds}$. Let $\sigma,\sigma'\in\cophs_{\Sigma}$. Combining the definition (\ref{eq:defpm}) with the bosonic Axiom (T3x) yields,
\begin{align*}
& A_{\hat{\Sigma}}\left(\tau^*(\iota_{\Sigma}^*(\sigma)\tens\sigma')\right)\\
& = \sum_{n,m} \overline{\rho_{\hat{\Sigma}}(\tau(\iota_{\Sigma}(\xi_{n})\tens\xi_{m}))}\,
\rho_{\hat{\Sigma}}(\tau(\iota_{\Sigma}(\sigma\, \xi_{n})\tens\sigma'\,\xi_{m})) \\
& = \sum_{n,m} \overline{\langle \xi_{n},\xi_{m}\rangle_{\Sigma}}\,
\langle \sigma\, \xi_{n}, \sigma'\,\xi_{m}\rangle_{\Sigma} \\
& = \lhs \sigma,\sigma'\rhs_{\Sigma} .
\end{align*}
Again, the resulting identity is analogous to Axiom (T3x) itself.

\subsection{Spaces on hypersurfaces}
\label{sec:bhsurfaces}

The considerations of the previous section suggest that we may formulate the core axioms directly with probability maps and do away with the amplitude maps altogether. This would replace Axioms (T4), (T3x), (T5a) and (T5b).

At the same time the probability maps are defined directly on operator spaces rather than the underlying Hilbert spaces. This suggests to eliminate the mention of the Hilbert spaces as well and replace the axioms referring to them by axioms referring to operator spaces. This would affect Axioms (T1), (T1b), (T2), (T2b). There are a priori different possibilities for what class of operators we should consider here. Close analogy with the original core axioms, especially with respect to Axioms (T5b) and (T3x), points towards the Hilbert-Schmidt operators with their inner product (\ref{eq:hsip}).

There is an intriguing consequence of such a step. Recall that the space $\cophs_{\Sigma}$ of Hilbert-Schmidt operators on the Hilbert space $\cH_{\Sigma}$ is canonically isomorphic to the tensor product of Hilbert spaces $\cH_{\Sigma}\ctens\cH_{\Sigma}^*$. But $\cH_{\Sigma}^*$ is canonically isomorphic to $\cH_{\overline{\Sigma}}$. Indeed, given $\psi\in\cH_{\overline{\Sigma}}$ the corresponding element in $\cH_{\Sigma}^*$ is the map $\eta\mapsto \langle\iota_{\overline{\Sigma}}(\psi),\eta\rangle_{\Sigma}$. So $\cophs_{\Sigma}$ is canonically isomorphic to $\cH_{\Sigma}\ctens\cH_{\overline{\Sigma}}$. This in turn is canonically isomorphic to $\cH_{\overline{\Sigma}}\ctens\cH_{\Sigma}$ by transposition. That is, the spaces of objects associated to the hypersurface $\Sigma$ for its two orientations are canonically isomorphic. This suggests to axiomatically associate just one single object to a hypersurface, irrespective of its orientation. We shall continue to refer to this object as $\cophs_{\Sigma}$. Given a region $M$, we also obtain a natural definition for the subspace $\cophsd_{\partial M}\subseteq \cophs_{\partial M}$ where the probability map $A_M$ will be well defined. This is, $\cophsd_{\partial M}\defeq \cH_{\Sigma}^\ds\tens\cH_{\overline{\Sigma}}^\ds$. This subspace is dense in $\cophs_{\partial M}$ with respect to the Hilbert-Schmidt inner product. Note that the definition of $A_M$ given in (\ref{eq:defpm}) reads in terms of the tensor product,
\begin{equation}
 A_M(\psi\tens\eta)=\rho_M(\psi)\overline{\rho_M(\iota_{\overline{\partial M}}(\eta))} .
\label{eq:probmaptp}
\end{equation}

Orientation change is still associated with a non-trivial operation on $\cophs_{\Sigma}$. Recall from Section~\ref{sec:bcomp} that this is the map $\iota_{\Sigma}^*:\cophs_{\Sigma}\to\cophs_{\overline{\Sigma}}$ given by $\sigma\mapsto\iota_{\Sigma}\circ\sigma\circ\iota_{\overline{\Sigma}}$. Identifying $\cophs_{\overline{\Sigma}}$ with $\cophs_{\Sigma}$ as indicated above shows that this is exactly taking the adjoint. That is, $\iota_{\Sigma}^*(\sigma)=\sigma^{\dagger}$. In terms of the tensor product presentation of $\cophs_{\Sigma}$ this is,
\begin{equation}
\iota_{\Sigma}^*(\psi\tens\eta)=\iota_{\overline{\Sigma}}(\eta)\tens \iota_{\Sigma}(\psi) .
\label{eq:iotastp}
\end{equation}
This may be seen as another hint that we should really consider a real formalism. That is, instead of considering the spaces $\cophs_{\Sigma}$ we restrict to the real subspaces $\cophsr_{\Sigma}$ of self-adjoint operators. As mentioned previously, $\cophsr_{\Sigma}$ is a real Hilbert space and $\cophs_{\Sigma}$ as a complex Hilbert space can be recovered as its complexification. Similarly, $A_M$ on $\cophsd_{\partial M}$ is recovered as the complex linear extension of $A_M$ on $\cophsrd_{\partial M}$.
With the restriction to $\cophsr_{\Sigma}$ the map $\iota_{\Sigma}^*$ becomes the identity. We may thus eliminate it altogether in the real formalism. As a consequence, the analogues of Axioms (T1b) and (T2b) disappear.

Our considerations so far also represent a transition from an oriented formalism to an unoriented formalism. To make this complete for regions we need to impose a consistency condition. If, in the spacetime system underlying the theory in question, the same region may appear with opposite orientations, the associated amplitudes need to be related. Given such an oriented region $M$ we denote its orientation reversed copy by $\overline{M}$. In order for the probability maps $A_M$ and $A_{\overline{M}}$ to yield the same physical probabilities we need the underlying amplitude maps to be related by,
\begin{equation}
 \rho_{\overline{M}}(\eta)=\overline{\rho_{M}(\iota_{\overline{\partial M}}(\eta))} .
\label{eq:amplor}
\end{equation}
This can be deduced easily from (\ref{eq:probmaptp}) which in fact may then be written in the symmetrical form,
\begin{equation}
 A_M(\psi\tens\eta)=\rho_M(\psi)\rho_{\overline{M}}(\eta) .
\label{eq:probmapsym}
\end{equation}

A structure on $\cophs_{\Sigma}$ that is lost in the restriction to the real subspace $\cophsr_{\Sigma}$ is the operator product. This is a problem since positivity plays an essential role in the probability interpretation (recall Section~\ref{sec:bprob}). But the identification of positive operators is accomplished with the operator product. The inner product structure on $\cophsr_{\Sigma}$ is not enough. On the other hand, not all of the product structure $\cophs_{\Sigma}$ seems to be operationally needed. A minimal proposal that we shall adopt in the following is to equip $\cophsr_{\Sigma}$ with the cone $\cophsp_{\Sigma}$ of positive elements, making it into an \emph{ordered vector space}. This allows to axiomatically implement positivity making the real formalism into a \emph{positive formalism}. Note that the order structure is compatible with the inner product of $\cophsr_{\Sigma}$ in the sense that
\begin{equation}
\lhs \sigma_1,\sigma_2\rhs_{\Sigma}\ge 0
\end{equation}
if $\sigma_1,\sigma_2\in \cophsp_{\Sigma}$. The order structure has further particular properties that we might want to enforce axiomatically. Basic properties include the facts that $\cophsp_{\Sigma}$ is a generating proper cone and that the order structure is Archimedean.



\subsection{A first axiomatization}
\label{sec:bpaxioms}

We present here explicitly an axiomatization of the positive formalism discussed in Sections~\ref{sec:bcomp} and \ref{sec:bhsurfaces}. The space previously identified as $\cophsr_{\Sigma}$ is here denoted $\cDr_{\Sigma}$ and analogously for the positive cone. To emphasize the parallelism to the bosonic core axioms (Appendix~\ref{sec:coreaxioms}) we keep the numbering, but we use the letter ``P'' (for ``positive'') instead of ``T''.

\begin{itemize}
\item[(P1)] Associated to each hypersurface $\Sigma$, irrespective of its orientation, is a real separable Hilbert space $\cDr_\Sigma$. We denote its inner product by $\lhs\cdot,\cdot\rhs_\Sigma$. Moreover, $\cDr_\Sigma$ is an Archimedean ordered vector space with generating proper cone $\cDp_\Sigma$ such that $\lhs \sigma,\sigma'\rhs_\Sigma\ge 0$ if $\sigma,\sigma'\in \cDp_\Sigma$.
\item[(P2)] Suppose the hypersurface $\Sigma$ decomposes into a disjoint
  union of hypersurfaces $\Sigma=\Sigma_1\cup\cdots\cup\Sigma_n$. Then,
  there is a positive isometric isomorphism of Hilbert spaces
  $\tau^*_{\Sigma_1,\dots,\Sigma_n;\Sigma}:\cDr_{\Sigma_1}\ctens\cdots\ctens\cDr_{\Sigma_n}\to\cDr_\Sigma$. The maps $\tau^*$ satisfy obvious associativity conditions.
\item[(P4)] Associated to each region $M$, irrespective of its orientation, is a positive linear map
  from a dense subspace $\cDrd_{\partial M}$ of $\cDr_{\partial M}$ to the real
  numbers, $A_M:\cDrd_{\partial M}\to\R$. This is called the
  \emph{probability map}.
\item[(P3x)] Let $\Sigma$ be a hypersurface. The boundary $\partial\hat{\Sigma}$ of the associated slice region $\hat{\Sigma}$ decomposes into the disjoint union $\partial\hat{\Sigma}=\overline{\Sigma}\cup\Sigma'$, where $\Sigma'$ denotes a second copy of $\Sigma$. Then, $\tau^*_{\overline{\Sigma},\Sigma';\partial\hat{\Sigma}}(\cDr_{\overline{\Sigma}}\tens\cDr_{\Sigma'})\subseteq\cDrd_{\partial\hat{\Sigma}}$. Moreover, $A_{\hat{\Sigma}}\circ\tau_{\overline{\Sigma},\Sigma';\partial\hat{\Sigma}}:\cDr_{\overline{\Sigma}}\tens\cDr_{\Sigma'}\to\R$ restricts to the inner product $\lhs\cdot,\cdot\rhs_\Sigma:\cDr_{\Sigma}\times\cDr_{\Sigma}\to\R$.
\item[(P5a)] Let $M_1$ and $M_2$ be regions and $M\defeq M_1\cup M_2$ be their disjoint union. Then $\partial M=\partial M_1\cup \partial M_2$ is also a disjoint union and $\tau^*_{\partial M_1,\partial M_2;\partial M}(\cDrd_{\partial M_1}\tens \cDrd_{\partial M_2})\subseteq \cDrd_{\partial M}$. Moreover, for all $\sigma_1\in\cDrd_{\partial M_1}$ and $\sigma_2\in\cDrd_{\partial M_2}$,
\begin{equation}
 A_{M}\left(\tau^*_{\partial M_1,\partial M_2;\partial M}(\sigma_1\tens\sigma_2)\right)= A_{M_1}(\sigma_1) A_{M_2}(\sigma_2) .
\end{equation}
\item[(P5b)] Let $M$ be a region with its boundary decomposing as a disjoint union $\partial M=\Sigma_1\cup\Sigma\cup \overline{\Sigma'}$, where $\Sigma'$ is a copy of $\Sigma$. Let $M_1$ denote the gluing of $M$ with itself along $\Sigma,\overline{\Sigma'}$ and suppose that $M_1$ is a region. Note $\partial M_1=\Sigma_1$. Then, $\tau^*_{\Sigma_1,\Sigma,\overline{\Sigma'};\partial M}(\sigma\tens\xi\tens\xi)\in\cDrd_{\partial M}$ for all $\sigma\in\cDrd_{\partial M_1}$ and $\xi\in\cDrd_\Sigma$. Moreover, for any orthonormal basis $\{\xi_i\}_{i\in I}$ of $\cDr_\Sigma$ in $\cDrd_{\Sigma}$, we have for all $\sigma\in\cDrd_{\partial M_1}$,
\begin{equation}
 A_{M_1}(\sigma)\cdot |c|^2(M;\Sigma,\overline{\Sigma'})
 =\sum_{i\in I} A_M\left(\tau^*_{\Sigma_1,\Sigma,\overline{\Sigma'};\partial M}(\sigma\tens\xi_i\tens\xi_i)\right),
\label{eq:glueax1}
\end{equation}
where $|c|^2(M;\Sigma,\overline{\Sigma'})\in\R^+$ is called the (modulus square of the) \emph{gluing anomaly factor} and depends only on the geometric data.
\end{itemize}


\subsection{Observables}
\label{sec:bobs}

In view of the considerations of Section~\ref{sec:bprob} it is rather straightforward to adapt the axioms for observables of \cite{Oe:obsgbf,Oe:feynobs} (see Appendix~\ref{sec:obsax}) to the positive formalism.

The way the composition of the expectation maps is induced by the composition of the observable maps is analogous to case of the probability maps. In particular, for the disjoint gluing of regions this can be read off by replacing in (\ref{eq:t5atrans}) the amplitude maps by observable maps. This yields Axiom (E2a) below from Axiom (O2a) in Appendix~\ref{sec:obsax}. For the gluing along hypersurfaces this arises by replacing in (\ref{eq:glueprob1}) the second occurrence of the amplitude map in each line with a corresponding observable map. This yields Axiom (E2b) below from Axiom (O2b) in Appendix~\ref{sec:obsax}.

\begin{itemize}
\item[(E1)] Associated to each spacetime region $M$ is a real vector space $\pobs_M$ of linear maps $\cDrd_{\partial M}\to\C$, called \emph{expectation maps}. In particular, $A_M\in\pobs_M$.
\item[(E2a)] Let $M_1$ and $M_2$ be regions and $M=M_1\cup M_2$ be their disjoint union. Then, there is an injective bilinear map $\aglue:\pobs_{M_1}\times\pobs_{M_2}\toi\pobs_{M}$ such that for all $A^{O_1}_{M_1}\in\pobs_{M_1}$ and $A^{O_2}_{M_2}\in\pobs_{M_2}$ and $\sigma_1\in\cDrd_{\partial M_1}$ and $\sigma_2\in\cDrd_{\partial M_2}$,
\begin{equation}
 A^{O_1}_{M_1}\aglue A^{O_2}_{M_2}(\tau^*_{\Sigma_1\cup\Sigma_2}(\sigma_1\tens\sigma_2))= A^{O_{1}}_{M_1}(\sigma_1) A^{O_{2}}_{M_2}(\sigma_2) .
\end{equation}
This operation is required to be associative in the obvious way.
\item[(E2b)] Let $M$ be a region with its boundary decomposing as a disjoint union $\partial M=\Sigma_1\cup\Sigma\cup \overline{\Sigma'}$ and $M_1$ given as in (P5b). Then, there is a linear map $\aglue_{\Sigma}:\pobs_{M}\to\pobs_{M_1}$ such that for all $A^O_M\in\pobs_{M}$ and any orthonormal basis $\{\xi_i\}_{i\in I}$ of $\cDr_\Sigma$ in $\cDrd_{\Sigma}$ and for all $\sigma\in\cDrd_{\partial M_1}$,
\begin{equation}
 \aglue_{\Sigma}(A^O_M)(\sigma)\cdot |c|^2(M;\Sigma,\overline{\Sigma'})
 =\sum_{i\in I}A^O_M(\tau^*_{\Sigma_1,\Sigma,\overline{\Sigma'};\partial M}(\sigma\tens\xi_i\tens\xi_i)) .
\end{equation}
This operation is required to commute with itself and with (E2a) in the obvious way.
\end{itemize}

We may also impose the condition
\begin{equation}
 \rho^O_{\overline{M}}(\psi)=\overline{\rho^O_M(\iota_{\partial M}(\psi))} ,
\label{eq:obsmor}
\end{equation}
analogous to the condition (\ref{eq:amplor}) for amplitudes. It induces on expectation maps the relation
\begin{equation}
 A^O_{\overline{M}}(\sigma)=\overline{A^O_M(\iota^*_{\partial M}(\sigma))} ,
\label{eq:expmor}
\end{equation}
which might be added to the axioms. We assume this in the following.

It might seem surprising that we do not restrict expectation maps to be real valued, even though they are defined here on the real vector space $\cDrd_{\partial M}$. In quantization schemes adapted to the standard formulation real classical observables are usually required to be represented by hermitian operators. These give rise to real expectation values. However, in quantum field theory, vacuum expectation values of real (time-ordered) observables are generically not real. (The prime example are n-point functions of a real field.) This has to do with the fact that observables in quantum field theory are spacetime objects. (See \cite{Oe:feynobs} for a discussion of this from the GBF perspective.) Thus, even if we restrict to observables that in some classical sense are real valued, we may not restrict expectation values to be real. For the special case of boundary observables the situation is different and more similar to the situation in the standard formulation. However, we shall not consider here a separate axiomatization for this case.
A consequence of the fact that expectation maps are not necessarily real is a resulting orientation dependence, as can be read off from equation (\ref{eq:expmor}). Even if $\sigma$ is self-adjoint, orientation change of the underlying region $M$ induces a complex conjugation of the expectation map.



\section{Fermionic and mixed theory}
\label{sec:fermionic}

We proceed to consider the more general case of general boundary quantum field theory with fermionic or mixed statistics.

In a quantum theory with fermionic degrees of freedom, the objects associated to hypersurfaces are graded Krein spaces rather than Hilbert spaces \cite{Oe:freefermi}. That is, the space $\cH_{\Sigma}$ associated with the hypersurface $\Sigma$ is a particular type of indefinite inner product space. $\cH_{\Sigma}$ decomposes into a direct sum of a \emph{positive part} $\cH_{\Sigma,+}$ where the inner product is positive definite,  and a \emph{negative part} $\cH_{\Sigma,-}$ where the inner product is negative definite.\footnote{As in \cite{Oe:freefermi} we use here a strict notion of Krein space with a fixed canonical decomposition into positive and negative parts. In contrast to \cite{Oe:freefermi} we indicate positive and negative parts here with lower indices.} $\cH_{\Sigma}$ is also a topological vector space in such a way that changing the signature of the inner product on its negative part makes it into a Hilbert space. We may think of the decomposition as a $\Z_2$-grading with $\cH_{\Sigma,+}$ the degree $0$ part and $\cH_{\Sigma,-}$ the degree $1$ part. We refer to this grading also as the \emph{signature}. $\cH_{\Sigma}$ also carries the usual $\Z_2$-grading that distinguishes even and odd fermion number. We refer to this as the fermionic grading or for short \emph{f-grading}. Both gradings are compatible in the sense that the positive and negative parts are individually f-graded.

Given $\psi\in\cH_{\Sigma}$ we denote by $\fdg{\psi}\in\{0,1\}$ its f-degree and by $\sig{\psi}\in\{0,1\}$ its signature. We also define the signature-detecting map $I:\cH_{\Sigma}\to\cH_{\Sigma}$ by
\begin{equation}
 I\, \psi \defeq (-1)^{\sig{\psi}} \psi .
\end{equation}

\subsection{Probabilities}
\label{sec:fprob}

As in the purely bosonic case we denote by $\cop_{\Sigma}$ the algebra of continuous operators on $\cH_{\Sigma}$.
The probability interpretation in the general case is essentially the same as in the purely bosonic case, except for two additional superselection rules \cite[Section~11]{Oe:freefermi}. The first of these is the usual f-grading of amplitudes, manifest in Axiom (T4) of Appendix~\ref{sec:coreaxioms}. That is, the amplitude map $\rho_M$ vanishes on the f-degree odd part of $\cH_{\partial M}$. To reflect this in the measurement process, the subspaces $\cA$ and $\cS$ determining ``knowledge'' and ``question'' should be taken to be subspaces of $\cH_{\partial M,0}$, the f-degree even part of $\cH_{\partial M}$. The other superselection rule imposes compatibility with the signature grading. That is, the subspaces $\cA$ and $\cS$ should decompose into direct sums $\cA_+\oplus\cA_-$ and $\cS_+\oplus\cS_-$ under signature. With these superselection rules in place we choose orthonormal basis as in the bosonic case and obtain the probability $P(\cA|\cS)$ via formula (\ref{eq:aprob}).

Taking the projection operator point of view as in formula (\ref{eq:aprobl}), the first superselection rule has the effect of restricting $\po_{\cA}$ and $\po_{\cS}$ to be part of the subalgebra of operators that are maps from $\cH_{\partial M,0}$ to itself. We denote this subalgebra by $\cop_{\partial M,00}$. The second superselection rule has the effect of further restricting $\po_{\cA}$ and $\po_{\cS}$ to also preserve signature. We denote the corresponding subalgebra by $\cop_{\partial M,00,+}\subseteq \cop_{\partial M,00}$.
The definition of the probability map (\ref{eq:defpm}) makes it manifestly dependent only on the subspace $\cop_{\partial M,00}$, since the amplitude map vanishes on $\cH_{\partial M,1}$.


\subsection{Real and positive structures}
\label{sec:frps}

In the purely bosonic case the Hilbert spaces $\cH_{\Sigma}$ and $\cH_{\overline{\Sigma}}$ associated with a hypersurface $\Sigma$ and its orientation reversed version are naturally conjugate linear isomorphic. As a consequence the spaces of operators on these, $\cop_{\Sigma}$ and $\cop_{\overline{\Sigma}}$ are naturally isomorphic. In the fermionic or mixed case the situation is more complicated due to the gradings. Since we are aiming for an orientation neutral point of view it is useful to model the operator spaces through the tensor product $\cH_{\Sigma}\ctens\cH_{\overline{\Sigma}}$. This was also done in the purely bosonic case in Section~\ref{sec:bhsurfaces} where this corresponded to taking the Hilbert-Schmidt operators on $\cH_{\Sigma}$ or $\cH_{\overline{\Sigma}}$. As there, we shall use the notation $\cophs_{\Sigma}$ and denote the inner product by $\lhs\cdot,\cdot\rhs_{\Sigma}$. This inner product is the one arising from the tensor product of Krein spaces and makes $\cophs_{\Sigma}$ into a Krein space. Moreover, we set the subspace $\cophsd_{\Sigma}$ to be $\cH_{\Sigma}^{\ds}\tens\cH_{\overline{\Sigma}}^{\ds}$.

F-grading and signature induce $\Z_2\times\Z_2$-gradings on $\cophs_{\Sigma}$. We shall denote by $\cophs_{\Sigma, i j}$ the $(i,j)$-graded part of $\cophs_{\Sigma}$, i.e., the subspace $\cH_{\Sigma, i}\ctens\cH_{\overline{\Sigma}, j}$. Here we set $i,j\in \{0,1\}$ for f-degree and $i,j\in \{+,-\}$ for signature. We also consider the $\Z_2$-gradings obtained by combining the two $\Z_2$-components. That is, $\cophs_{\Sigma, i}$ denotes the direct sum of the spaces $\cophs_{\Sigma, j k}$ such that $j+k\equiv i$. This is in agreement with the notation already introduced in Section~\ref{sec:fprob}. Also, with this notation, the positive part of $\cophs_{\Sigma}$ is precisely $\cophs_{\Sigma,+}$ while the negative part is $\cophs_{\Sigma,-}$. $\cophs_{\Sigma}$ is isometrically isomorphic as a Krein space to $\cophs_{\overline{\Sigma}}$ via $\tau_{\overline{\Sigma},\Sigma;\Sigma\cup\overline{\Sigma}}^{-1}\circ\tau_{\Sigma,\overline{\Sigma};\Sigma\cup\overline{\Sigma}}$, which is the f-graded transposition (see Axiom (T2) in Appendix~\ref{sec:coreaxioms}). Note that the isometry interchanges the order of the graded components, i.e., $\cophs_{\Sigma, i j}$ is isomorphic to $\cophs_{\overline{\Sigma}, j i}$.

Going back to an interpretation of the elements of $\cophs_{\Sigma}$ as operators on $\cH_{\Sigma}$ or $\cH_{\overline{\Sigma}}$ is less straightforward than in the purely bosonic case. However, in view of the probability interpretation as outlined above in terms of operators a natural identification is given as follows. Let $\psi\in\cH_{\Sigma}$ and $\eta\in\cH_{\overline{\Sigma}}$. Then, $\psi\tens\eta$ acts as an operator on $\cH_{\Sigma}$ as follows,
\begin{equation}
 (\psi\tens\eta)\xi=\psi\,\langle I\, \iota_{\overline{\Sigma}}(\eta),\xi\rangle_{\Sigma} .
\end{equation}
The probability map $A_M:\cophsd_{\partial M}\to\C$ given by (\ref{eq:defpm}) then satisfies (\ref{eq:probmaptp}) as in the purely bosonic case. Moreover, as there we shall impose the orientation compatibility condition (\ref{eq:amplor}), leading to formula (\ref{eq:probmapsym}) for the probability map.

A natural notion of \emph{adjoint} or complex conjugation, i.e., \emph{real structure} on $\cophs_{\Sigma}$ is given, as in the purely bosonic case, by $\iota_{\Sigma}^*$ defined by formula (\ref{eq:iotastp}). This is just the composition of $\iota_{\Sigma\cup\overline{\Sigma}}$ with the relevant $\tau$-maps. We continue to use the notation $\sigma^{\dagger}=\iota_{\Sigma}^*(\sigma)$. Note, however, that in contrast to the purely bosonic case, the adjoint does not coincide in general with the map $\sigma\mapsto \iota_{\Sigma}\circ\sigma\circ\iota_{\overline{\Sigma}}$ for $\sigma\in \cophs_{\Sigma}$ viewed as an operator. From the operator point of view the adjoint is given by the formula,
\begin{equation}
\langle I \sigma^{\dagger} \xi,\eta\rangle_{\Sigma} = \langle I \xi,\sigma\eta\rangle_{\Sigma} .
\end{equation}
That is, the notion of adjoint here is precisely that coming from the Hilbert space structure of $\cH_{\Sigma}$, i.e., from $\cH_{\Sigma}$ viewed as the Hilbert space $\cH_{\Sigma,+}\oplus\overline{\cH_{\Sigma,-}}$. Here $\overline{\cH_{\Sigma,-}}$ is the same inner product space as $\cH_{\Sigma,-}$, except for a reversal of the sign of the inner product. This same identification of $\cH_{\Sigma}$ with a Hilbert space may be used to define the notion of a \emph{positive} element of $\cophs_{\Sigma}$. We denote in the following by $\cophsr_{\Sigma}$ and $\cophsp_{\Sigma}$ the subsets of self-adjoint and of positive elements respectively. This leads to the correct notion in terms of the probability interpretation (recall Section~\ref{sec:fprob}), although there a notion of positivity is only needed in the subspace $\cophs_{\Sigma,00,+}\subseteq\cophs_{\Sigma}$, due to the superselection rules. In particular, the probability map $A_M$ is real on $\cophsrd_{\partial M}$ and positive on $\cophspd_{\partial M}$.

The inner product of $\cophs_{\Sigma}$ in terms of the operator point of view may be written as,
\begin{equation}
\lhs\sigma',\sigma\rhs_{\Sigma}
=\sum_{n\in N} (-1)^{\fdg{\zeta_n}+\sig{\zeta_n}} \langle \sigma'\zeta_n,\sigma\zeta_n\rangle_{\Sigma} ,
\end{equation}
where $\{\zeta_n\}_{n\in N}$ is an orthonormal basis of $\cH_{\Sigma}$, generalizing (\ref{eq:hsip}). Note that the inner product is compatible with the real structure in an f-graded sense (compare \cite{Oe:freefermi}),
\begin{equation}
 \lhs\sigma_1^\dagger,\sigma_2^\dagger\rhs_{\Sigma}
 =(-1)^{\fdg{\sigma}}\overline{\lhs\sigma_1,\sigma_2\rhs_{\Sigma}} .
\end{equation}
(Here $\fdg{\sigma}$ stands for the equivalent choices $\fdg{\sigma_1}$, $\fdg{\sigma_2}$ or $\fdg{\sigma_1}\fdg{\sigma_2}$.) In particular, the inner product is real on $\cophsr_{\Sigma,0}$ and imaginary on $\cophsr_{\Sigma,1}$. Also, it is positive on the set $\cophspd_{\Sigma,0,+}$.


\subsection{Composition}
\label{sec:fcomp}

We consider in this section the transfer of the composition properties from amplitude maps to probability maps for the general case, generalizing the considerations in Section~\ref{sec:bcomp}. We limit ourselves to highlighting the differences to the purely bosonic case.

In order to handle the decomposition of hypersurfaces we need analogues of the $\tau$-maps for the spaces $\cophs_{\Sigma}$. These are simply assembled from the usual $\tau$-maps by using the definition of $\cophs_{\Sigma}$ in terms of the tensor product $\cH_{\Sigma}\ctens\cH_{\overline{\Sigma}}$. For a hypersurface decomposition $\Sigma=\Sigma_1\cup\cdots\cup\Sigma_n$ we denote the associated map $\cophs_{\Sigma_1}\ctens\cdots\ctens\cophs_{\Sigma_n}\to\cophs_{\Sigma}$ by $\tau^*_{\Sigma_1,\dots,\Sigma_n;\Sigma}$. The associativity of the $\tau$-maps makes these well defined and also associative. Note that this definition coincides with (\ref{eq:taubos}) in the purely bosonic case, but not in general.

Concerning the gluing Axiom (T5a) of Appendix~\ref{sec:coreaxioms} for the gluing of disjoint regions, the identity
\begin{equation}
A_M(\tau^*(\sigma_1\tens\sigma_2))=A_{M_1}(\sigma_1) A_{M_2}(\sigma_2)
\end{equation}
is established precisely as in the purely bosonic case, compare (\ref{eq:t5atrans}) in Section~\ref{sec:bcomp}. We abbreviate here $\tau^*_{\partial M_1,\partial M_2;\partial M}$ by $\tau^*$. Note that due to the f-bigrading of $A_M$ we may take $\sigma_1$ and $\sigma_2$ to lie in $\cophsd_{\partial M_1,00}$ and $\cophsd_{\partial M_2,00}$ respectively which implies $\tau^*(\sigma_1\tens\sigma_2)\in\cophsd_{\partial M,00}$. As in the purely bosonic case realness or positivity of $\sigma_1$ and $\sigma_2$ then imply the same property for $\tau^*(\sigma_1\tens\sigma_2)$.

We proceed to consider the gluing along hypersurfaces. In contrast to the treatment of the purely bosonic case in Section~\ref{sec:bcomp} we use the tensor product point of view rather than the operator point of view on $\cophs_{\Sigma}$. Assume the context of Axiom (T5b). To simplify notation we leave out the explicit mention of the $\tau$-maps. Note that the condition (\ref{eq:amplor}) implies for the gluing anomaly,
\begin{equation}
 c(\overline{M};\overline{\Sigma},\Sigma')
=\overline{c(M;\Sigma,\overline{\Sigma'})} .
\end{equation}
Let $\{\zeta_m\}_{m\in N}$ be an orthonormal basis of $\cH_{\Sigma}$ in $\cH_{\Sigma}^{\ds}$. Let $\psi\in\cH^{\ds}_{\partial M_1}$ and $\eta\in\cH^{\ds}_{\overline{\partial M_1}}$.
Combining the equality (\ref{eq:probmapsym}) and the gluing identity (\ref{eq:glueid5b}) yields,
\begin{align}
& A_{M_1}(\psi\tens\eta) \cdot |c|^2 \nonumber\\
& = \rho_{M_1}(\psi)\cdot c\;\rho_{\overline{M_1}}(\eta)\cdot \overline{c} \nonumber\\
& = \sum_{k,l\in N} (-1)^{\sig{\zeta_k}+\sig{\iota_{\Sigma}(\zeta_l)}}\rho_M\left(\psi\tens\zeta_k\tens\iota_{\Sigma}(\zeta_k)\right)\;
\rho_{\overline{M}}\left(\eta\tens\iota_{\Sigma}(\zeta_l)\tens\zeta_l\right) \nonumber\\
& = \sum_{k,l\in N} (-1)^{\sig{\zeta_k}+\sig{\iota_{\Sigma}(\zeta_l)}}
 A_M\left((\psi\tens\zeta_k\tens\iota_{\Sigma}(\zeta_k))\tens(\eta \tens \iota_{\Sigma}(\zeta_l) \tens \zeta_l)\right) \nonumber\\
& = \sum_{k,l\in N} (-1)^{\sig{\zeta_k\tens\iota_{\Sigma}(\zeta_l)}}
 A_M\left((\psi\tens\eta)\tens(\zeta_k\tens \iota_{\Sigma}(\zeta_l))\tens \iota_{\Sigma\cup\overline{\Sigma'}}(\zeta_k \tens \iota_{\Sigma}(\zeta_l))\right)  .
\end{align}
We note that $\{\zeta_k\tens \iota_{\Sigma}(\zeta_l)\}_{(k,l)\in N\times N}$ is an orthonormal basis of $\cophs_{\Sigma}$. Indeed, choosing an arbitrary orthonormal basis $\{\xi_{n}\}_{n\in N}$ of $\cophs_{\Sigma}$ we might write the identity as,
\begin{equation}
A_{M_1}(\sigma) \cdot |c|^2
= \sum_{n\in N} (-1)^{\sig{\xi_{n}}} A_M\left(\tau^*_{\Sigma_1,\Sigma,\overline{\Sigma'},\partial M}(\sigma\tens \xi_n\tens \iota_{\Sigma}^*(\xi_n))\right) ,
\end{equation}
generalizing (\ref{eq:bglueid5b}).

Finally we consider Axiom (T3x) relating amplitude map and inner product. Thus, given a hypersurface $\Sigma$ we let $\psi,\psi'\in\cH_{\Sigma}$ and $\eta,\eta'\in\cH_{\overline{\Sigma}}$. Omitting again the $\tau$- and $\tau^*$-maps we have from identity (\ref{eq:probmapsym}) and Axiom (T3x),
\begin{align}
& A_{\hat{\Sigma}}(\iota^*_{\Sigma}(\psi'\tens\eta') \tens (\psi\tens\eta)) \nonumber \\
& = A_{\hat{\Sigma}}(\iota_{\Sigma}(\psi')\tens\psi\tens\iota_{\overline{\Sigma}}(\eta')\tens\eta) \nonumber \\
& = \rho_{\hat{\Sigma}}(\iota_{\Sigma}(\psi')\tens\psi)\,
 \rho_{\overline{\hat{\Sigma}}}(\iota_{\overline{\Sigma}}(\eta')\tens\eta) \nonumber \\
& = \langle \psi',\psi\rangle_{\Sigma} \langle \eta',\eta\rangle_{\overline{\Sigma}} \nonumber \\
& = \lhs \psi'\tens\eta',\psi\tens\eta\rhs_{\Sigma} .
\end{align}
Note that the absence of sign factors from the gradings is due to the fact that the amplitude map vanishes on the f-degree odd subspace.



\subsection{A first axiomatization}
\label{sec:fpaxioms}

We present in this section an axiomatization of the positive formalism in the general case. Apart from the additional structure coming from the gradings and worked out in the previous sections there is another crucial difference for the axiomatization. In the purely bosonic case it was possible to formulate the axioms at the level of the real spaces $\cophsr_{\Sigma}$ instead of the complex spaces $\cophs_{\Sigma}$. This is not possible in the general case. The reason for this is that the the tensor product does not commute with the real structure in a simple way, but in an f-graded way. That is, leaving out $\tau$-maps, we have for elements $\sigma_1\in\cophs_{\Sigma_1}$, $\sigma_2\in\cophs_{\Sigma_2}$,
\begin{equation}
 (\sigma_1\tens\sigma_2)^{\dagger}=\sigma_2^{\dagger}\tens\sigma_1^{\dagger}
 =(-1)^{\fdg{\sigma_1}\cdot\fdg{\sigma_2}} \sigma_1^{\dagger}\tens\sigma_2^{\dagger}.
\end{equation}
In particular, the tensor product of self-adjoint elements is not necessarily self-adjoint. This implies that we need to retain complex spaces and analogues of Axioms (T1b) and (T2b) in spite of the fact that $\cophs_{\Sigma}$ is canonically isomorphic to $\cophs_{\overline{\Sigma}}$ and that $A_M$ on $\cophsd_{\Sigma}$ can be recovered completely from $A_M$ on $\cophsrd_{\Sigma}$.

Since the axioms do not seem to require bi-gradings and using simple gradings makes them simpler, we only use simple gradings in the axioms. This relaxes the condition that the probability map $A_M$ is non-vanishing only on $\cophs_{\partial M,00}$ to the weaker condition that it may be non-vanishing on $\cophs_{\partial M,0}$. Whether this might be physically justified or have physical significance is unclear at the moment. As in the purely bosonic case we mark the axioms with the letter ``P'' for ``positive''. We denote the analogues of the spaces $\cophs_{\Sigma}$ by $\cD_{\Sigma}$.

\begin{itemize}
\item[(P1)] Associated to each hypersurface $\Sigma$, irrespective of its orientation, is a complex separable f-graded Krein space $\cD_\Sigma$. We denote its inner product by $\lhs\cdot,\cdot\rhs_\Sigma$. Moreover, there is a distinguished real vector space $\cDr_\Sigma$, such that $\cD_\Sigma$ is the complexification of $\cDr_\Sigma$. $\cDr_\Sigma$ is an Archimedean ordered vector space with generating proper cone $\cDp_\Sigma$ such that $\lhs \sigma,\sigma'\rhs_\Sigma\ge 0$ if $\sigma,\sigma'\in \cDp_{\Sigma,0,+}$.
\item[(P1b)] Associated to each hypersurface $\Sigma$ is a conjugate linear adapted f-graded isometric involution $\iota^*_{\Sigma}:\cD_{\Sigma}\to\cD_{\Sigma}$, providing a real structure. In particular, $\cDr_\Sigma$ is the real subspace of $\cD_\Sigma$ invariant under $\iota^*_{\Sigma}$.
\item[(P2)] Suppose the hypersurface $\Sigma$ decomposes into a disjoint
  union of hypersurfaces $\Sigma=\Sigma_1\cup\cdots\cup\Sigma_n$. Then,
  there is an isometric isomorphism of Krein spaces
  $\tau^*_{\Sigma_1,\dots,\Sigma_n;\Sigma}:\cD_{\Sigma_1}\ctens\cdots\ctens\cD_{\Sigma_n}\to\cD_\Sigma$, positive on the real restriction $\cDr_{\Sigma_1,0}\ctens\cdots\ctens\cDr_{\Sigma_n,0}\to\cDr_{\Sigma,0}$. The maps $\tau^*$ satisfy obvious associativity conditions. Moreover, in the case $n=2$ the map $(\tau^*_{\Sigma_2,\Sigma_1;\Sigma})^{-1}\circ \tau^*_{\Sigma_1,\Sigma_2;\Sigma}:\cD_{\Sigma_1}\ctens\cD_{\Sigma_2}\to\cD_{\Sigma_2}\ctens\cD_{\Sigma_1}$ is the f-graded transposition,
\begin{equation}
 \sigma_1\tens\sigma_2\mapsto (-1)^{\fdg{\sigma_1}\cdot\fdg{\sigma_2}}\sigma_2\tens\sigma_1 .
\end{equation}
\item[(P2b)] Real structure and decomposition are compatible in an f-graded sense. That is, for a disjoint decomposition of hypersurfaces $\Sigma=\Sigma_1\cup\Sigma_2$ we have
\begin{equation}
\tau^*_{\Sigma_1,\Sigma_2;\Sigma}
 \left(\iota^*_{\Sigma_1}(\sigma_1)\tens\iota^*_{\Sigma_2}(\sigma_2)\right)
  =(-1)^{\fdg{\sigma_1}\cdot\fdg{\sigma_2}}\iota^*_\Sigma\left(\tau^*_{\Sigma_1,\Sigma_2;\Sigma}(\sigma_1\tens\sigma_2)\right) .
\end{equation}
\item[(P4)] Associated to each region $M$ is an f-graded linear map
  from a dense subspace $\cDd_{\partial M}$ of $\cD_{\partial M}$ to the complex
  numbers, $A_M:\cDd_{\partial M}\to\C$. This is called the
  \emph{probability map}. Moreover, $A_M$ is positive on $\cDpd_{\partial M,0}$ (and thus real on $\cDrd_{\partial M,0}$). Also, if $\overline{M}$ is a region, $A_{\overline{M}}$ and $A_M$ are related via,
\begin{equation}
 A_{\overline{M}}(\sigma)=\overline{A_M(\iota^*_{\partial M}(\sigma))} .
\end{equation}
\item[(P3x)] Let $\Sigma$ be a hypersurface. The boundary $\partial\hat{\Sigma}$ of the associated slice region $\hat{\Sigma}$ decomposes into the disjoint union $\partial\hat{\Sigma}=\overline{\Sigma}\cup\Sigma'$, where $\Sigma'$ denotes a second copy of $\Sigma$. Then, $\tau^*_{\overline{\Sigma},\Sigma';\partial\hat{\Sigma}}(\cD_{\overline{\Sigma}}\tens\cD_{\Sigma'})\subseteq\cDd_{\partial\hat{\Sigma}}$. Moreover, $A_{\hat{\Sigma}}\circ\tau^*_{\overline{\Sigma},\Sigma';\partial\hat{\Sigma}}:\cD_{\overline{\Sigma}}\tens\cD_{\Sigma'}\to\C$ restricts to the pairing $\lhs\iota^*_{\overline{\Sigma}}(\cdot),\cdot\rhs_\Sigma:\cD_{\overline{\Sigma}}\times\cD_{\Sigma}\to\C$.
\item[(P5a)] Let $M_1$ and $M_2$ be regions and $M\defeq M_1\cup M_2$ be their disjoint union. Then $\partial M=\partial M_1\cup \partial M_2$ is also a disjoint union and $\tau^*_{\partial M_1,\partial M_2;\partial M}(\cDd_{\partial M_1}\tens \cDd_{\partial M_2})\subseteq \cDd_{\partial M}$. Moreover, for all $\sigma_1\in\cDd_{\partial M_1}$ and $\sigma_2\in\cDd_{\partial M_2}$,
\begin{equation}
 A_{M}\left(\tau^*_{\partial M_1,\partial M_2;\partial M}(\sigma_1\tens\sigma_2)\right)= A_{M_1}(\sigma_1) A_{M_2}(\sigma_2) .
\end{equation}
\item[(P5b)] Let $M$ be a region with its boundary decomposing as a disjoint union $\partial M=\Sigma_1\cup\Sigma\cup \overline{\Sigma'}$, where $\Sigma'$ is a copy of $\Sigma$. Let $M_1$ denote the gluing of $M$ with itself along $\Sigma,\overline{\Sigma'}$ and suppose that $M_1$ is a region. Note $\partial M_1=\Sigma_1$. Then, $\tau^*_{\Sigma_1,\Sigma,\overline{\Sigma'};\partial M}(\sigma\tens\xi\tens\iota^*_{\Sigma}(\xi))\in\cDd_{\partial M}$ for all $\sigma\in\cDd_{\partial M_1}$ and $\xi\in\cDd_\Sigma$. Moreover, for any orthonormal basis $\{\xi_i\}_{i\in I}$ of $\cD_\Sigma$ in $\cDd_{\Sigma}$, we have for all $\sigma\in\cDd_{\partial M_1}$,
\begin{equation}
 A_{M_1}(\sigma)\cdot |c|^2(M;\Sigma,\overline{\Sigma'})
 =\sum_{i\in I} (-1)^{\sig{\xi_i}} A_M\left(\tau^*_{\Sigma_1,\Sigma,\overline{\Sigma'};\partial M}(\sigma\tens\xi_i\tens\iota^*_{\Sigma}(\xi_i))\right),
\end{equation}
where $|c|^2(M;\Sigma,\overline{\Sigma'})\in\R^+$ is called the (modulus square of the) \emph{gluing anomaly factor} and depends only on the geometric data.
\end{itemize}


\subsection{Observables}
\label{sec:fobs}

Observables in the GBF in the presence of fermionic degrees have not been properly discussed before. Before moving to the positive formalism we shall thus spend some time to discuss them here.

We recall first basic aspects of the standard formulation in this respect. There, the definition of the expectation value of an observable in a state does not depend on the presence or not of fermionic degrees of freedom. However, the presence of fermionic degrees of freedom comes with an associated superselection rule. That is, the state in question is required to have (in our language) a definite f-degree. This means that the expectation value will vanish if the observable itself is f-degree odd. Indeed, $n$-point functions with odd $n$ for a fermionic field vanish in quantum field theory. On the other hand, observables with odd f-degree are still important as they may be composed to form observables of even f-degree. Indeed, the standard field operators for fermionic fields in quantum field theory have odd f-degree.

Generalizing to the GBF changes little. As in the purely bosonic case the objects that encode observables in a region $M$ are the observable maps $\rho^O_M:\cH_{\partial M}\to\C$. The additional structure in the presence of fermionic degrees of freedom is the f-grading on $\rho^O_M:\cH^{\ds}_{\partial M}\to\C$. Concretely, $\rho^O_M$ has even f-degree if it vanishes on $\cH^{\ds}_{\partial M,1}$ and odd f-degree if it vanishes on $\cH^{\ds}_{\partial M,0}$. The discussion of expectation values of observables in the GBF based on formula (\ref{eq:expval}) is essentially the same as that given in Section~\ref{sec:bprob}, except for the crucial addition of the superselection rules as discussed in Section~\ref{sec:fprob}. The upshot is that bosonic observables (that is observables with even f-degree) can be treated as in a purely bosonic theory, while the fermionic observables (that is observables with odd f-degree) have vanishing expectation values. The latter is easy to read off from equation (\ref{eq:expval}) taking into account that the superselection rules force the amplitude map as well as the projector $\po_{\cS}$ to have even f-degree. Note also that the recovery of expectation values of the standard formulation from the GBF ones, demonstrated in \cite{Oe:obsgbf}, remains valid for bosonic observables even in the presence of fermionic degrees of freedom.

The axiomatic treatment of observable maps can be carried out almost exactly as in the purely bosonic case, see Appendix~\ref{sec:obsax}. The only visible modification is the inclusion of a signature factor in relation (\ref{eq:obsglueid}) of Axiom (O2b). Note also that there is no explicit mention of an f-grading of the observable maps in the axioms. This merely means that (in contrast to the amplitude maps) there is no restriction for them to be bosonic. There are striking consequences of the presence of fermionic degrees of freedom nevertheless. In particular, the gluing operation in Axiom (O2a) is no longer commutative, but becomes order-dependent. That is,
\begin{equation}
\rho^{O_1}_{M_1}\aglue \rho^{O_2}_{M_2}=(-1)^{\fdg{\rho^{O_1}_{M_1}}\cdot\fdg{\rho^{O_2}_{M_2}}} \rho^{O_2}_{M_2}\aglue \rho^{O_1}_{M_1} .
\label{eq:ordobsmcomp}
\end{equation}

We proceed to consider expectation maps and their axiomatization. We use the very same definition (\ref{eq:defexpm}) of expectation map as in the purely bosonic case. In terms of the tensor product point of view the expectation map thus takes the form,
\begin{equation}
 A^O_M(\psi\tens\eta)=\rho^O_M(\psi) \overline{\rho_{M}(\iota_{\overline{\partial M}}(\eta))} .
\end{equation}
The way the composition axioms of observable maps induce composition axioms for expectation maps generalizes straightforwardly from the purely bosonic case. As with the core axioms an important distinction to the treatment of the purely bosonic case is the use of the complex Krein spaces $\cD_{\Sigma}$ instead of their real counterparts $\cDr_{\Sigma}$. Another difference is the additional signature factor in Axiom (E2b), analogous to the difference in Axiom (P5b). For fermionic observables an order-dependence of the composition of expectation maps is induced from the order-dependence (\ref{eq:ordobsmcomp}) of the underlying composition of observable maps. Thus, we have,
\begin{equation}
A^{O_1}_{M_1}\aglue A^{O_2}_{M_2}=(-1)^{\fdg{A^{O_1}_{M_1}}\cdot\fdg{A^{O_2}_{M_2}}}A^{O_2}_{M_2}\aglue A^{O_1}_{M_1} .
\end{equation}

\begin{itemize}
\item[(E1)] Associated to each spacetime region $M$ is a real vector space $\pobs_M$ of linear maps $\cDd_{\partial M}\to\C$, called \emph{expectation maps}. In particular, $A_M\in\pobs_M$.
\item[(E2a)] Let $M_1$ and $M_2$ be regions and $M=M_1\cup M_2$ be their disjoint union. Then, there is an injective bilinear map $\aglue:\pobs_{M_1}\times\pobs_{M_2}\toi\pobs_{M}$ such that for all $A^{O_1}_{M_1}\in\pobs_{M_1}$ and $A^{O_2}_{M_2}\in\pobs_{M_2}$ and $\sigma_1\in\cDd_{\partial M_1}$ and $\sigma_2\in\cDd_{\partial M_2}$,
\begin{equation}
 A^{O_1}_{M_1}\aglue A^{O_2}_{M_2}(\tau^*_{\Sigma_1\cup\Sigma_2}(\sigma_1\tens\sigma_2))= A^{O_{1}}_{M_1}(\sigma_1) A^{O_{2}}_{M_2}(\sigma_2) .
\end{equation}
This operation is required to be associative in the obvious way.
\item[(E2b)] Let $M$ be a region with its boundary decomposing as a disjoint union $\partial M=\Sigma_1\cup\Sigma\cup \overline{\Sigma'}$ and $M_1$ given as in (P5b). Then, there is a linear map $\aglue_{\Sigma}:\pobs_{M}\to\pobs_{M_1}$ such that for all $A^O_{M}\in\pobs_{M}$ and any orthonormal basis $\{\xi_i\}_{i\in I}$ of $\cD_\Sigma$ in $\cDd_{\Sigma}$ and for all $\sigma\in\cDd_{\partial M_1}$,
\begin{equation}
 \aglue_{\Sigma}(A^O_M)(\sigma)\cdot |c|^2(M;\Sigma,\overline{\Sigma'})
 =\sum_{i\in I} (-1)^{\sig{\xi_i}} A^O_M(\tau^*_{\Sigma_1,\Sigma,\overline{\Sigma'};\partial M}(\sigma\tens\xi_i\tens\iota^*_{\Sigma}(\xi_i))) .
\end{equation}
This operation is required to commute with itself and with (E2a) in the obvious way.
\end{itemize}

As in the purely bosonic case we may wish to impose the orientation compatibility condition for observable maps (\ref{eq:obsmor}). From the tensor product point of view the expectation map then takes the form,
\begin{equation}
 A^O_M(\psi\tens\eta)=\rho^O_M(\psi)\rho_{\overline{M}}(\eta) .
\end{equation}
The resulting relation for expectation maps is (\ref{eq:expmor}), which again, we might want to incorporate into the axioms.



\section{Discussion and outlook}
\label{sec:outlook}

We have presented in this work a new formalism for encoding quantum theories in the general boundary formulation (GBF). The extraction of predicted measurement probabilities and expectation values is more direct in this formalism. In particular, the positivity of probabilities is directly imprinted in terms of order structure in the formalism, whence we term it the \emph{positive formalism}. From an operational point of view it has less ``excess baggage'' than the usual formalism that we shall refer to as the \emph{amplitude formalism}.

The transition from the amplitude formalism to the positive formalism in the GBF is somewhat analogous to the transition from a pure state formalism to a mixed state formalism in the standard formulation. In both cases a Hilbert (or Krein) space is replaced by a space of suitable operators on it. There are crucial differences, however, both physically and (in consequence) mathematically. The elements of the Hilbert space in the standard formulation are interpreted as states, that is as determining a possible reality of the system as a whole. This interpretation is in general not tenable for the elements of a Hilbert (or Krein) space associated with a hypersurface in the GBF. Rather than the elements it is the projectors of such a Hilbert (or Krein) space that are used to extract physical information (see Section~\ref{sec:bprob}). These projectors may be thought of as encoding the \emph{quantum boundary conditions} of the measurement. Elements of these spaces are special only in so far as they represent one-dimensional projectors and are thus something like \emph{elementary} quantum boundary conditions. The positive formalism reflects the operational relevance of the projectors by associating spaces to hypersurfaces that directly contain them. In contrast to the spaces of mixed states in the standard formulation, elements of these spaces cannot be interpreted in general as encoding ensembles of systems. They can be interpreted, however, as encoding ensembles of quantum boundary conditions for the question asked in a measurement (see Section~\ref{sec:bprob}).

Note also that a state that is mixed in terms of the standard formulation may not necessarily appear ``mixed'' in the GBF. Indeed, the first explicit use of a mixed state in the GBF occurred in an investigation of the Unruh effect in terms of the GBF \cite{CoRa:unruh}. It turns out there that the mixed state induced by the Minkowski vacuum in Rindler spacetime is representable essentially as an ordinary vector in the relevant boundary Hilbert space, rather than as a density operator.

In spite of the conceptual differences, we recall that the standard formulation is reproducible from the GBF when the former makes sense. Suppose thus a globally hyperbolic background spacetime and consider spacelike Cauchy hypersurfaces with a global time orientation. In the amplitude formalism the Hilbert spaces associated to these hypersurfaces provide then ``copies'' of ``the'' Hilbert space of the standard formulation. The amplitude maps for regions bounded by pairs of these hypersurfaces recover the standard transition amplitudes. In the positive formalism the positive elements of the ordered vector space associated to a Cauchy hypersurface provide a copy of the standard space of (unnormalized) mixed states. The probability maps yield the transition amplitudes for these mixed states. The usual normalization of mixed states in terms of the trace can be recovered when sufficient structure is given on the ordered vector spaces. However, while this normalization is essential in the standard formulation to preserve probability, it is less relevant in the positive formalism due to the quotient structure of probability expressions.

The elaboration of the positive formalism as presented in this work should be seen as merely a first draft. In particular, choices made in the axiomatization (Sections~\ref{sec:bpaxioms} and \ref{sec:fpaxioms}) should be seen as provisional. In line with the operational motivation for the introduction of the positive formalism, we should throw away as much information as possible in the transition, without diminishing the physical interpretability. This motivated us (in Section~\ref{sec:bhsurfaces}) to ``forget'' in the axiomatization the algebra structure on the spaces $\cD_{\Sigma}$ coming from thinking of them as spaces of operators. While we did retain the order structure, it is not quite clear whether it is sufficient to identify the operationally relevant projectors and their pertinent properties. One possibility for retaining more structure while not keeping the full operator product would be to keep the Jordan algebra structure. That is, we would equip the real vector space of self-adjoint operators with the Jordan product,
\begin{equation}
 \sigma_1\bullet\sigma_2\defeq \frac{1}{2}(\sigma_1\sigma_2+\sigma_2\sigma_1) .
\end{equation}
This would be particularly suggestive in the purely bosonic case, where the axiomatization (Section~\ref{sec:bpaxioms}) is precisely in terms of the spaces of self-adjoint operators.
Even with respect to the structure of $\cDr_{\Sigma}$ as an ordered vector space we might want to require further properties. For example, the space of self-adjoint operators on a Hilbert space is special from an order perspective in that it forms an anti-lattice \cite{Kad:orderpropbsaop}. This suggests to require that $\cDr_{\Sigma}$ be an anti-lattice.

Related to questions of structure of the spaces $\cD_{\Sigma}$ is the question of their ``size''. In particular, we have chosen spaces of Hilbert-Schmidt operators to provide the blueprint for the spaces $\cD_{\Sigma}$ associated to hypersurfaces $\Sigma$. A disadvantage of this choice is the lack of infinite-dimensional projectors in these spaces. In particular, the projectors $\po_{\cS}$, representing the ``preparation'' in a quantum measurement (recall Section~\ref{sec:bprob}) are typically infinite-dimensional. Thus, to really evaluate formula (\ref{eq:nprob}) or (\ref{eq:nexpval}) we typically need to insert suitable sums or limits. Enlarging the spaces $\cD_{\Sigma}$, e.g., to correspond to all continuous operators would clash with the Hilbert-Schmidt inner product on $\cD_{\Sigma}$. However, we might choose not to view this inner product as a primary structure on $\cD_{\Sigma}$. Indeed, Axiom (P3x) (see Sections~\ref{sec:bpaxioms} and \ref{sec:fpaxioms}) suggests that we may view it instead as induced from the probability map. Taking this seriously, we may remove the explicit imposition of the inner product and do away with Axiom (P3x) as well, replacing it perhaps merely with some non-degeneracy condition. The inner product would still be there, arising in the style of Axiom (P3x), but perhaps defined on a subspace of $\cD_{\Sigma}$ only. This also opens the way for considering other topologies on $\cD_{\Sigma}$. To this end we remark the following: The order topology on the space of continuous operators on a Hilbert space viewed merely as an ordered vector space recovers precisely its usual operator norm topology. What is more, with the unit element considered as an order unit we even recover exactly the operator norm (see e.g.\ \cite{Sch:tvs}). Another aspect of the ``size'' of $\cD_{\Sigma}$ is that the amplitude map cannot be defined on the whole space but only on a subspace, even if we take as the blueprint for $\cD_{\Sigma}$ only the Hilbert-Schmidt operators. This problem is of course inherited from the amplitude formalism, but positivity suggests a solution as indicated in Section~\ref{sec:probprop}: Restricting the amplitude map to the positive elements while extending its range to include $\infty$.

The positive formalism is presented here in such a way that any quantum theory obeying the axioms of the usual amplitude formalism can be converted straightforwardly to a quantum theory obeying the axioms of the positive formalism. The reverse is not the case, and indeed should not be the case if we have succeeded to any degree in our goal of eliminating operationally irrelevant information. To convert a theory obeying the axioms of the positive formalism to one obeying the axioms of the amplitude formalism additional structure has to be added. There might be a natural choice for this structure, there might be many choices or there might be none at all. The freedom gained in not needing this operationally irrelevant structure for constructing quantum theories is a key advantage of the positive formalism, more important in our opinion than possible gains in elegance and naturalness. We expect in particular that this could help in the quest of finding a truly local description of quantum field theory, i.e., a description where not only observables but also ``states'' are localized.

Another advantage of the positive formalism over the amplitude formalism is its potential to interconnect with methods of quantum information theory. In particular, we may consider maps that are analogous to expectation maps, but that are not induced from conventional observables as in formula (\ref{eq:defexpm}). Rather they may implement more general quantum operations, in which case we shall refer to them here as \emph{operation maps}.
We notice here a certain convergence of the presented formalism with the operator tensor formulation of quantum theory proposed by L.~Hardy \cite{Har:optensqt}. To see this consider a spacetime region with its boundary decomposed into various components. We divide the boundary components into incoming and outgoing ones, depending on some global directional information (e.g., an arrow of time). The operation map for the spacetime region can be converted to an operator from the tensor product of the incoming spaces to the tensor product of the outgoing spaces. These are then analogous to the operators representing quantum operations in the operator tensor formulation. What is more, given a number of quantum operations in adjacent spacetime regions, the gluing of the regions yields a contraction of the associated operation maps according to Axioms (E2a) and (E2b). This is analogous to the contraction of the corresponding operators in the operator tensor formulation.
There remain key differences between the operator tensor formulation and the GBF, however. One of these is the time asymmetry of the former which ultimately comes from its conceptual reliance on the standard formulation of quantum theory. Overcoming this would presumably require a generalization of the probability interpretation underlying the operator tensor formulation in the sense of the GBF.

From a more technical point of view, purely as an axiomatic system of topological quantum field theory, the positive formalism exhibits interesting differences to the amplitude formalism. The most striking one is perhaps that the latter is oriented while the former is unoriented, at least in the purely bosonic case (Section~\ref{sec:bpaxioms}). In the fermionic or mixed case there are ``residual'' orientation dependencies, in terms of complex conjugations, due to the order (and thus orientation) dependence of the graded tensor product. In any case, the transition from the amplitude formalism to the positive formalism is suggestive of a more general procedure for topological quantum field theories of constructing the ``modulus squared'' theory of a given theory.


\subsection*{Acknowledgments}

I would like to thank Lucien Hardy for convincing me, while visiting him at the Perimeter Institute in Waterloo, Canada in January 2007, of the possibility and advantage of doing quantum theory in a formalism ``linear'' in probabilities. The positive formalism described in this work grew out of this inspiration.
This work was supported in part by UNAM--DGAPA--PAPIIT through project grant IN100212.

\appendix

\section{GBF axioms in the amplitude formalism}
\label{sec:gbfaxioms}

\subsection{Spacetime system}
\label{sec:sts}

We recall aspects of the way spacetime is encoded in the GBF abstractly through a \emph{spacetime system} \cite{Oe:gbqft,Oe:freefermi}. The latter consists of:
\begin{itemize}
\item A collection of oriented topological manifolds of dimension $d$ with boundary and possibly with additional structure. These are called \emph{regions}.
\item A collection of oriented topological manifolds of dimension $d-1$ without boundary and possibly with additional structure. These are called \emph{hypersurfaces}.
\end{itemize}
These collections satisfy additional properties:
\begin{itemize}
\item Regions and hypersurfaces may only have a finite number of connected components.
\item The boundary of a region is a hypersurface.
\item Every connected component of a region is a region and every connected component of a hypersurface is a hypersurface.
\item There is a notion of \emph{decomposition} of a hypersurface which consists in presenting the hypersurface as a disjoint union of other hypersurfaces.
\item There is a notion of \emph{gluing} of regions which consists in presenting a region as the union of regions such that the interiors are disjoint and the intersection of the original regions is a hypersurface.
\end{itemize}
There is also a modified notion of region, called a \emph{slice region}.\footnote{The term slice region was used for the first time in \cite{Oe:freefermi}. In previous papers this was called an ``empty region''.} This is really a hypersurface $\Sigma$ that is treated as if it was a region, denoted $\hat{\Sigma}$ with boundary $\Sigma\cup\overline{\Sigma}$.

\subsection{Core axioms}
\label{sec:coreaxioms}

The core axioms of the GBF are presented here in the form given in \cite{Oe:freefermi}. In contrast to earlier version, this permits the inclusion of fermionic degrees of freedom. The latter come with a $\Z_2$-grading, the fermionic or \emph{f-grading}.

\begin{itemize}
\item[(T1)] Associated to each hypersurface $\Sigma$ is a complex
  separable f-graded Krein space $\cH_\Sigma$. We denote its indefinite inner product by
  $\langle\cdot,\cdot\rangle_\Sigma$.
\item[(T1b)] Associated to each hypersurface $\Sigma$ is a conjugate linear adapted f-graded isometry $\iota_\Sigma:\cH_\Sigma\to\cH_{\overline{\Sigma}}$. This map is an involution in the sense that $\iota_{\overline{\Sigma}}\circ\iota_\Sigma$ is the identity on  $\cH_\Sigma$.
\item[(T2)] Suppose the hypersurface $\Sigma$ decomposes into a disjoint
  union of hypersurfaces $\Sigma=\Sigma_1\cup\cdots\cup\Sigma_n$. Then,
  there is an isometric isomorphism of Krein spaces
  $\tau_{\Sigma_1,\dots,\Sigma_n;\Sigma}:\cH_{\Sigma_1}\ctens\cdots\ctens\cH_{\Sigma_n}\to\cH_\Sigma$. The maps $\tau$ satisfy obvious associativity conditions. Moreover, in the case $n=2$ the map $\tau_{\Sigma_2,\Sigma_1;\Sigma}^{-1}\circ \tau_{\Sigma_1,\Sigma_2;\Sigma}:\cH_{\Sigma_1}\ctens\cH_{\Sigma_2}\to\cH_{\Sigma_2}\ctens\cH_{\Sigma_1}$ is the f-graded transposition,
\begin{equation}
 \psi_1\tens\psi_2\mapsto (-1)^{\fdg{\psi_1}\cdot\fdg{\psi_2}}\psi_2\tens\psi_1 .
\end{equation}
\item[(T2b)] Orientation change and decomposition are compatible in an f-graded sense. That is, for a disjoint decomposition of hypersurfaces $\Sigma=\Sigma_1\cup\Sigma_2$ we have
\begin{equation}
\tau_{\overline{\Sigma}_1,\overline{\Sigma}_2;\overline{\Sigma}}
 \left(\iota_{\Sigma_1}(\psi_1)\tens\iota_{\Sigma_2}(\psi_2)\right)
  =(-1)^{\fdg{\psi_1}\cdot\fdg{\psi_2}}\iota_\Sigma\left(\tau_{\Sigma_1,\Sigma_2;\Sigma}(\psi_1\tens\psi_2)\right) .
\end{equation}
\item[(T4)] Associated to each region $M$ is a f-graded linear map
  from a dense subspace $\cH_{\partial M}^\ds$ of
  $\cH_{\partial M}$ to the complex
  numbers, $\rho_M:\cH_{\partial M}^\ds\to\C$. Here $\partial M$ denotes the boundary of $M$ with the induced orientation. This is called the
  \emph{amplitude} map.
\item[(T3x)] Let $\Sigma$ be a hypersurface. The boundary $\partial\hat{\Sigma}$ of the associated empty region $\hat{\Sigma}$ decomposes into the disjoint union $\partial\hat{\Sigma}=\overline{\Sigma}\cup\Sigma'$, where $\Sigma'$ denotes a second copy of $\Sigma$. Then, $\rho_{\hat{\Sigma}}$ is well defined on $\tau_{\overline{\Sigma},\Sigma';\partial\hat{\Sigma}}(\cH_{\overline{\Sigma}}\tens\cH_{\Sigma'})\subseteq\cH_{\partial\hat{\Sigma}}$. Moreover, $\rho_{\hat{\Sigma}}\circ\tau_{\overline{\Sigma},\Sigma';\partial\hat{\Sigma}}$ restricts to a bilinear pairing $(\cdot,\cdot)_\Sigma:\cH_{\overline{\Sigma}}\times\cH_{\Sigma'}\to\C$ such that $\langle\cdot,\cdot\rangle_\Sigma=(\iota_\Sigma(\cdot),\cdot)_\Sigma$.
\item[(T5a)] Let $M_1$ and $M_2$ be regions and $M\defeq M_1\cup M_2$ be their disjoint union. Then $\partial M=\partial M_1\cup \partial M_2$ is also a disjoint union and $\tau_{\partial M_1,\partial M_2;\partial M}(\cH_{\partial M_1}^\ds\tens \cH_{\partial M_2}^\ds)\subseteq \cH_{\partial M}^\ds$. Moreover, for all $\psi_1\in\cH_{\partial M_1}^\ds$ and $\psi_2\in\cH_{\partial M_2}^\ds$,
\begin{equation}
 \rho_{M}\left(\tau_{\partial M_1,\partial M_2;\partial M}(\psi_1\tens\psi_2)\right)= \rho_{M_1}(\psi_1)\rho_{M_2}(\psi_2) .
\label{eq:glueid5a}
\end{equation}
\item[(T5b)] Let $M$ be a region with its boundary decomposing as a disjoint union $\partial M=\Sigma_1\cup\Sigma\cup \overline{\Sigma'}$, where $\Sigma'$ is a copy of $\Sigma$. Let $M_1$ denote the gluing of $M$ with itself along $\Sigma,\overline{\Sigma'}$ and suppose that $M_1$ is a region. Note $\partial M_1=\Sigma_1$. Then, $\tau_{\Sigma_1,\Sigma,\overline{\Sigma'};\partial M}(\psi\tens\xi\tens\iota_\Sigma(\xi))\in\cH_{\partial M}^\ds$ for all $\psi\in\cH_{\partial M_1}^\ds$ and $\xi\in\cH_\Sigma$. Moreover, for any orthonormal basis $\{\zeta_i\}_{i\in I}$ of $\cH_\Sigma$ in $\cH^{\ds}_{\Sigma}$, we have for all $\psi\in\cH_{\partial M_1}^\ds$,
\begin{equation}
 \rho_{M_1}(\psi)\cdot c(M;\Sigma,\overline{\Sigma'})
 =\sum_{i\in I}(-1)^{\sig{\zeta_i}}\rho_M\left(\tau_{\Sigma_1,\Sigma,\overline{\Sigma'};\partial M}(\psi\tens\zeta_i\tens\iota_\Sigma(\zeta_i))\right),
\label{eq:glueid5b}
\end{equation}
where $c(M;\Sigma,\overline{\Sigma'})\in\C\setminus\{0\}$ is called the \emph{gluing anomaly factor} and depends only on the geometric data.
\end{itemize}


\subsection{Observable axioms}
\label{sec:obsax}

Observables were introduced into the GBF and axiomatized in \cite{Oe:obsgbf}. We extend here the axiomatization in the form presented in \cite{Oe:feynobs} from the purely bosonic case to the general case.\footnote{Instead of denoting observables by letters such as $O$ as in the axioms presented in \cite{Oe:obsgbf} and \cite{Oe:feynobs} we use a notation of the type $\rho^O_M$. This facilitates the identification with expectation maps $A^O_M$ encoding the same observable.}

\begin{itemize}
\item[(O1)] Associated to each spacetime region $M$ is a real vector space $\obs_M$ of linear maps $\cH_{\partial M}^\ds\to\C$, called \emph{observable maps}. In particular, $\rho_M\in\obs_M$.
\item[(O2a)] Let $M_1$ and $M_2$ be regions and $M=M_1\cup M_2$ be their disjoint union. Then, there is an injective bilinear map $\aglue:\obs_{M_1}\times\obs_{M_2}\toi\obs_{M}$ such that for all $\rho^{O_1}_{M_1}\in\obs_{M_1}$ and $\rho^{O_2}_{M_2}\in\obs_{M_2}$ and $\psi_1\in\cH_{\partial M_1}^\ds$ and $\psi_2\in\cH_{\partial M_2}^\ds$,
\begin{equation}
 A^{O_1}_{M_1}\aglue A^{O_2}_{M_2}(\psi_1\tens\psi_2)
 = A^{O_{1}}_{M_1}(\psi_1) A^{O_{2}}_{M_2}(\psi_2) .
\end{equation}
This operation is required to be associative in the obvious way.
\item[(O2b)] Let $M$ be a region with its boundary decomposing as a disjoint union $\partial M=\Sigma_1\cup\Sigma\cup \overline{\Sigma'}$ and $M_1$ given as in (T5b). Then, there is a linear map $\aglue_{\Sigma}:\obs_{M}\to\obs_{M_1}$ such that for all $A^O_M\in\obs_{M}$ and any orthonormal basis $\{\xi_i\}_{i\in I}$ of $\cH_\Sigma$ in $\cH_{\Sigma}^{\ds}$ and for all $\psi\in\cH_{\partial M_1}^\ds$,
\begin{equation}
 \aglue_{\Sigma}(A^O_M)(\psi)\cdot c(M;\Sigma,\overline{\Sigma'})
 =\sum_{i\in I} (-1)^{\sig{\xi_i}} A^O_M(\psi\tens\xi_i\tens\iota_\Sigma(\xi_i)) .
\label{eq:obsglueid}
\end{equation}
This operation is required to commute with itself and with (O2a) in the obvious way.
\end{itemize}

\bibliographystyle{stdnodoi} 
\bibliography{stdrefsb}
\end{document}